\documentclass[10pt,final,journal]{Template/IEEEtran}

\usepackage[cmex10]{amsmath}
\usepackage{mathtools}
\usepackage{psfrag}
\usepackage{multicol}
\usepackage{epstopdf}
\usepackage{graphicx,bm,upgreek,eucal}
\PassOptionsToPackage{bookmarks=false}{hyperref}
\usepackage{subfigure,textcomp,gensymb,url,multirow}
\usepackage[justification=centering]{caption}
\usepackage{color,cite}
\usepackage{algorithm}
\usepackage{algpseudocode}
\usepackage{palatino}
\usepackage{anysize, booktabs, setspace}
\marginsize{0.7in}{0.7in}{0.7in}{0.7in}
\usepackage{newpxmath}
\usepackage{amsfonts, amssymb}
\usepackage{cleveref, nomencl, blindtext, mathtools, etoolbox}
\usepackage[export]{adjustbox}

\newcommand{\comments}[1]{}
\newtheorem{theorem}{Theorem}
\newtheorem{lemma}{Lemma}
\allowdisplaybreaks
\usepackage{tikz}
\usetikzlibrary{shapes,arrows,spy,positioning,patterns,shapes.geometric}
\usetikzlibrary{external}
\makeatletter
\def\munderbar#1{\underline{\sbox\tw@{$#1$}\dp\tw@\z@\box\tw@}}
\makeatother
\newcommand{\floor}[1]{\lfloor #1 \rfloor}
\providecommand{\myceil}[1]{\lceil #1 \rceil}

\pdfoutput=1

\newcommand*\circled[1]{\tikz[baseline=(char.base)]{
            \node[shape=circle,draw,inner sep=2pt] (char) {#1};}}
\newtheorem{corollary}{Corollary}

\title{\LARGE \bf
Persistent Monitoring of Dynamically Changing Environments Using an Unmanned Vehicle
}

\author{
  \small
S. K. K. Hari$^{1}$, S. Rathinam$^{1}$, S. Darbha$^{1}$, K. Kalyanam$^{2}$, S. G. Manyam$^{2}$, D. Casbeer$^{2}$%
\thanks{$^{1}$ Department of Mechanical Engineering, Texas A\&M University, College Station, U.S.A. 77845.}%
\thanks{$^{2}$ 
Control Science Center of Excellence,
Autonomous Control Branch,
Aerospace Systems Directorate,
Wright-Patterson A.F.B., OH 45433.
}%
}

\begin{document}

\maketitle

\begin{abstract}
We consider the problem of planning a closed walk $\mathcal W$ for a UAV to persistently monitor a finite number of stationary targets with equal priorities and dynamically changing properties. A UAV must physically visit the targets in order to monitor them and collect information therein. The frequency of monitoring any given target is specified by a target revisit time, $i.e.$, the maximum allowable time between any two successive visits to the target. The problem considered in this paper is the following: Given $n$ targets and $k \geq n$ allowed visits to them, find an optimal closed walk $\mathcal W^*(k)$ so that every target is visited at least once and the maximum revisit time over all the targets, $\mathcal R(\mathcal W(k))$, is minimized. We prove the following: If $k \geq n^2-n$, $\mathcal R(\mathcal W^*(k))$ (or simply, $\mathcal R^*(k)$) takes only two values: $\mathcal R^*(n)$ when $k$ is an integral multiple of $n$, and $\mathcal R^*(n+1)$ otherwise. This result suggests significant computational savings -- one only needs to determine $\mathcal W^*(n)$ and $\mathcal W^*(n+1)$ to construct an optimal solution $\mathcal W^*(k)$. We provide MILP formulations for computing $\mathcal W^*(n)$ and $\mathcal W^*(n+1)$. Furthermore, for {\it any} given $k$, we prove that $\mathcal R^*(k) \geq \mathcal R^*(k+n)$. 
\end{abstract}


\section{Introduction}

Dynamically changing environments are typical in a wide variety of applications ranging from military to agricultural domains. In many applications, an up-to-date knowledge of the changing parameters is crucial \cite{casbeer2005forest, clark2005cooperative, rathinam2008vision} for managing uncertainty. Often, the environment to be monitored is vast, and the coverage area of sensors is limited. Since it is practically impossible to sweep the entire region with a sensor's footprint, the environment is usually divided into a set of disjointed regions represented by stationary targets, the properties of which are indicators of the properties of the environment.  Persistent monitoring of the environment or targets can be achieved using a set of vehicles equipped with appropriate sensors that travel across the region to collect the required information.

Monitoring a target requires a vehicle to travel to the target (referred to as a visit to the target), and collect information. The collected information is then sent to a base station, where it is analyzed by a human operator to take appropriate actions \cite{hari2016scheduling}. To ensure timely actions, maintaining freshness of information is pivotal. Persistent monitoring requires that the time between successive revisits to a target be as small as possible and that the data received by the operator is as fresh as possible. There are many problem determinants for persistent monitoring applications: for example, the number of vehicles, vehicle-target assignment constraints, communication capabilities of vehicles, persistent monitoring requirements of every target, fuel and refueling constraints. Given the problem determinants, the core problem of persistent monitoring is to find the sequence of targets to be visited by each vehicle so that the maximum revisit time between successive revisits to any target is minimized. 

In this study, we assume that the information collected at the targets is instantly transmitted to the base station. In doing so, we implicitly assume that the communication between a vehicle and its base station is unrestricted. When the base station is out of the communication range of the vehicle, information must be physically transported. In such cases, one must pay attention to preserving the freshness of information \cite{kalyanam2017average} in addition to the typical requirements of persistence and efficiency of monitoring, which are discussed in the following paragraphs.

Persistence is crucial in dynamic environments, as a stoppage in monitoring leads to a continuous growth of uncertainty in the properties of interest. Persistently monitoring a target requires it to be visited again and again. However, fuel limitations of a vehicle may impose additional constraints on persistent monitoring. For example, if the vehicle has limited fuel on-board, it may require refueling before proceeding with revisiting the set of targets again. In this case, not only is the fuel carried by the vehicle on-board important, but also the refueling time; this is because persistent monitoring is characterized by the maximum time between successive revisits to any target. If the vehicle can carry limited fuel on-board, then there is a frequent need to refuel; in this article, we assume that refueling is done at a depot.

Targets can be closely monitored with multiple vehicles; however, there are a number of issues that arise with multiple vehicles: for example, not all vehicles may have the sensor suite to monitor every target; in this case, vehicle-target assignment constraints must be obeyed. Vehicle-target assignment constraints also subsume the important issue of what constitutes a visit to a target -- should it be performed by the same vehicle or by different vehicles? Is there a precedence constraint for these visits? In this article, we sidestep these issues and consider the basic problem of persistent monitoring with a single vehicle; the results from the single vehicle problem can be used to build solutions for persistent monitoring with multiple vehicles.  We further assume that there is a single depot and it is a target where the vehicle starts from with the maximum possible fuel it can store and returns to for refueling. 

The everlasting nature of the persistent monitoring together with the limited fuel capacity of the vehicle, requires the vehicle to be periodically refueled \cite{nigam2012control}. In this work, we surrogate the vehicle's fuel capacity by the number of visits after which it must be refueled (say $k$ visits). After every $k$ visits, the vehicle is refueled at the depot, after which the persistent monitoring task is resumed. There are four other assumptions we make in this paper: (1) triangle inequality is satisfied by travel times between targets, i.e., it is faster for a vehicle to go directly from a target $A$ to a target $B$ than through some other intermediate target $C$. There is a tacit assumption that a vehicle can travel without any restraint from any target to any other target. (2) the endurance time of a vehicle (time taken by a vehicle to empty its fuel tank or discharge its battery from a fully charged condition) is proportional to the fuel capacity and that there is no loss of generality in surrogating the fuel capacity (fuel carried on-board by a vehicle) with the number of visits. If one can solve the problem of routing vehicles for persistent monitoring with a specified number, $k$, of visits, one can compute the fuel required, say $\phi_{fuel}(k)$; by triangle inequality, this function is monotonically increasing in $k$.  The main result of this paper can be used to compute this function; knowing $\phi_{fuel}$ one can find the number $k$ of visits for any given fuel capacity and number of targets to be visited. (3) Here, we also assume that the time taken to refuel/recharge the vehicle is negligible compared to the travel times between the targets; this is not an unreasonable assumption, especially when discharged battery packs can be removed and charged battery packs can be quickly loaded onto the vehicle and this time can be assumed negligible when compared to the travel time. (4)  From the perspectives of convenience and efficiency (as in battery recharging), there is a tacit assumption made: a vehicle can be refueled only when $k$ visits have been made (i.e., fuel/charge has been expended fully before recharging).

The problem considered in this paper is the following:
  Given that  there are $n$ targets and the vehicle starts at the depot with a fuel capacity equivalent to $k\ge n$ visits, find a route of $k$ visits so that the vehicle (1) visits each target at least once, and return to the depot for its $k^{th}$ visit and (2) the maximum time between successive revisits to any target is minimized. Note that the vehicle refuels after the $k^{th}$ visit at the depot; since $k \geq n$, the vehicle is allowed to make multiple visits to targets. In graph-theoretic terminology, this route of $k$ visits is referred to as a {\it closed walk} $\mathcal W(k)$ (more about this will be discussed in Section \ref{sec:problem-statement}). To ensure persistence, this walk is repeated over and over. The maximum revisit time over all the targets, when the walk is repeated continuously, is referred to as the {\it revisit time of the walk} $\mathcal R(\mathcal W(k))$. The objective of this work is to identify a closed walk $\mathcal W^*(k)$ with $k$ visits such that the revisit time of the walk $\mathcal R(\mathcal W^*(k))$, is the minimum. For a given number of visits, the lowest possible revisit time is referred to as the {\it optimal revisit time} $\mathcal R^*(k)$, and a walk $\mathcal W^*(k)$ with the minimum revisit time (i.e., $\mathcal R(\mathcal W^*(k)) =\mathcal R^*(k)$) is referred to as an optimal walk. 


When the number of allowable visits $k$ is equal to the number of targets, each target is visited exactly once in any walk. In such a case, the revisit time of any target is equal to the time taken to complete a closed tour visiting all the targets once. Hence, finding a walk with the least maximum revisit time is equivalent to finding a traveling salesman tour over all the targets, which is very well known to be NP-Hard. Hence, the persistent monitoring problem is also NP-Hard. The computational difficulty of the problem increases with the number of visits.

The problem of persistent monitoring in the ideal case of unlimited fuel capacity takes the same form as the one considered here; this same problem was considered in \cite{magesh2011persistent}; in this work, heuristic algorithms were provided for the closed walk. A generalization of this problem was considered in \cite{smith}, where some targets are considered more important and  their revisit time is weighed differently; the corresponding objective of routing is to minimize the maximum weighted revisit time of targets. In \cite{smith}, the authors proved that an optimal infinite walk (a walk with infinite number of visits) can be constructed by concatenating a finite closed walk (a walk with the same initial and terminal node) with itself multiple times. Nonetheless, they showed that the size of the finite closed walk can be arbitrarily large. They proved that the problem is APX-Hard, and provided two polynomial time approximation algorithms, with approximation ratios $O(log \: n)$ and $O(log \: \rho)$, where $\rho$ is the ratio of maximum and minimum weights of the targets. In \cite{ccta2018}, the authors posed the problem as an MILP, and provided an iterative sub-optimal scheme for the problem. The sub-optimal scheme was observed to be an order of magnitude faster than the optimal scheme, without compromising much on the quality of solutions.

In \cite{chevaleyre2004theoretical}, the author considered the problem of multi-agent patrolling of equally weighted targets, with a focus on developing efficient team patrolling strategies, based on a TSP tour over the targets. A TSP tour does not allow multiple visits to targets in a cycle, as opposed to closed walks considered in the present article. The author proposed two team patrolling strategies: cyclic based strategy and partition-based strategy and compared their performance based on a metric of minimizing the team revisit time.
In \cite{pasqualetti2012cooperative}, the authors extended the problem to weighted targets (which was proved to be NP-hard in \cite{pasqualetti2012complexity}). The authors in \cite{pasqualetti2012cooperative} construct a non-intersecting tour (without repeated visits to targets) based on a minimum spanning tree over the targets, and compare two team patrolling strategies: equal spacing strategy, and equal time spacing strategy over the the tour. Additionally, they propose a distributed control algorithm to equally space robots along a shared trajectory, considering two different communication models (passing communication model and neighborhood-broadcast communication model). The authors of \cite{cannata2011minimalist} consider a relatively simpler problem of multi-robot frequency coverage, in which the number of visits made to each target is pre-specified. They propose an algorithm that does not require inter-robot communication, and has a low computational requirement. In contrast to the above articles, the authors in \cite{smith2012persistent} considered fixed trajectories, and developed velocity controllers for robots/agents to achieve minimum revisit time.

\subsection{Contributions}
In the present article, we consider the problem of monitoring equally weighted targets, and provide results characterizing the structure of optimal solutions. Given $k \geq n$ allowable visits for the vehicle, we prove the following:

\begin{enumerate}
    \item $\mathcal R^*(k) \geq TSP^*$, $ \forall  k \geq n$; equality holds when $k$ is an integral multiple of $n$.
    \item $\mathcal R^*(k)$ is bi-modal for $k \geq n^2-n$; it takes only two values, $\mathcal R^*(n)$ if $k$ is an integral multiple of $n$, and $\mathcal R^*(n+1)$ otherwise.
    \item $\mathcal R^*(k+n) \leq \mathcal R^*(k)$, $\forall k \geq n$.
    \item $\mathcal R^*(k)$ is a monotonically non-decreasing function of $k$, for $n \leq k\leq 2n-1$, .
    \item $\mathcal R^*(k) = \mathcal R^*(n+\myceil{\frac{q}{p}})$, where $k > n$ is not an integral multiple of $n$, and can be expressed in the quotient remainder form as $pn + q$, such that $p \geq 1$, $1 \leq q \leq n-1$ and $p, q \in Z_+$.

\end{enumerate}

These results indicate that one needs to solve at most $n$ problems (from $n$ visits to $2n-1$ visits) to completely determine optimal solutions for any given $k \geq n$ and can lead to a substantial reduction in the computational effort required to solve the problem for higher number of visits. 


\subsection{Organization}
The rest of the paper is organized as follows: The exact problem statement is presented in Section \ref{sec:problem-statement}. Section \ref{sec:proofs} provides useful lemmas and theorems that characterizes optimal solutions. Section \ref{sec:sim} provides numerical simulations to corroborate the results proved in Section \ref{sec:proofs}, using an MILP formulation presented in Appendix \ref{sec:form}. Conclusions and future directions are provided in Section \ref{sec:conc}.

\section{Problem Statement} \label{sec:problem-statement}

 Consider a set of $n$ spatially distributed targets/nodes that needs to be monitored by a UAV. Let the targets be denoted by the set $\mathcal{T}=\{1,2,\cdots,n\}$, and the travel times between any two distinct targets $u,v$ ($u \neq v$) in the set be denoted by $c(u,v)>0$. We assume that the targets are of equal priorities, and the travel times between them satisfy the triangle inequality, i.e., $c(u,v) + c(v,w)\geq c(u,w)$ for all $u,v,w \in \mathcal{T}$. \\ 

 The UAV tasked with monitoring the targets must be recharged after every $k$ visits. The UAV visits the targets in the order specified by the sequence $\mathcal W=(v_1,v_2,\cdots,v_{k+1})$ (such that $v_i\in \mathcal{T}$ for $i=1,\cdots,k+1$); this sequence is repeated after every $k$ visits. Here, we assume $k \geq n$, which allows each target to be visited at least once and some targets to be visited multiple times in a cycle of $k$ visits. Due to repetition of nodes/vertices, this sequence is referred to as a {\it walk}. The node $v_1$ is the first or starting node of the walk from which the vehicle starts its mission, and is not counted in the number of visits made to targets. The node $v_{k+1}$ is the last visited target in the walk.
 
 We assume that the vehicle starts from and returns to the depot (chosen as of one the targets), after every $k$ visits. Consequently, we have $v_1 = v_{k+1}$, and a walk with the same initial and terminal node is referred to as a  closed walk. Note that a visit is counted only if a vehicle travels between two {\it distinct} targets. Therefore, we have $v_i \neq v_{i+1}$ for $i=1,\cdots,k$. Moreover, the vehicle must visit all the targets in a walk of $k$ visits. In contrast with the graph-theoretic terminology\cite{papadimitriou1982combinatorial}, we use the term {\it closed walk}  differently to refer to a walk (a) with the same initial and terminal nodes, (b) in which every target is visited at least once and (c) successive visits correspond to distinct targets. Since we primarily deal with only closed walks in this paper, we use walks and closed walks interchangeably when the context is clear. 
 
\begin{figure}[h!]
    \centering
    \includegraphics[scale=0.45]{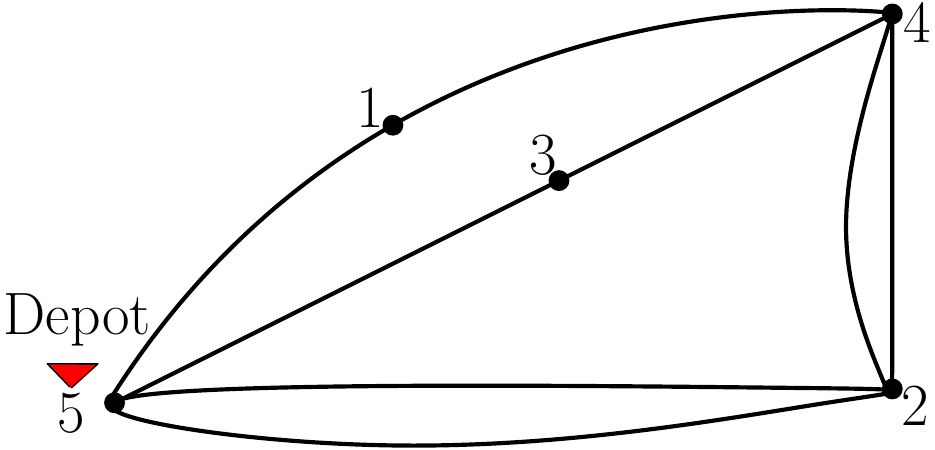}
    \caption{Graphical representation of a walk $\mathcal W(8) = (5,1,4,2,5,3,4,2,5)$ with 8 visits}
    \label{fig:walk}
\end{figure}

\begin{figure}[h!]
    \centering
   \includegraphics[scale=0.45]{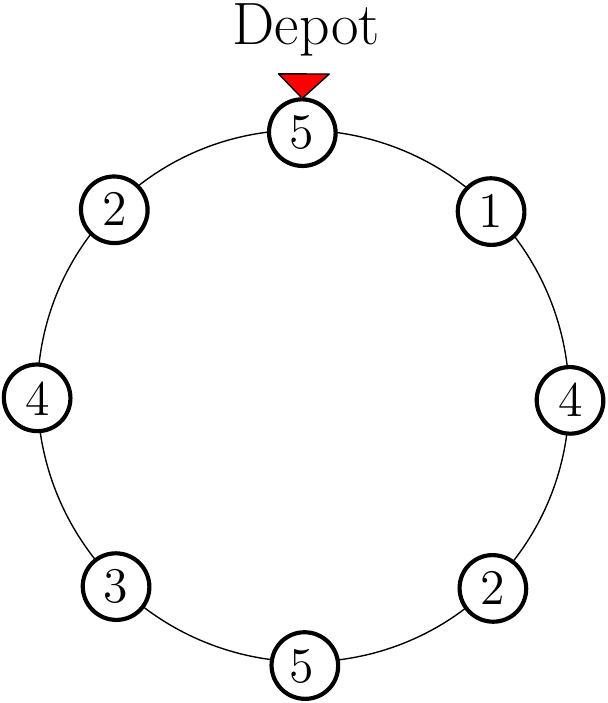}
    \caption{A cyclic representation of the walk $\mathcal W(8) = (5,1,4,2,5,3,4,2,5)$ (shown in Figure \ref{fig:walk}) with node 5 as the depot}
    \label{fig:cyclicrep}
\end{figure}

 For the purpose of illustration, consider Figure \ref{fig:walk}, which shows a closed walk of eight visits. Since a closed walk is continuously repeated after every $k$ visits (in Figure \ref{fig:walk}, $k=8$), it is often depicted using a cyclic representation as shown in Figure \ref{fig:cyclicrep}. The total time required to traverse through all the nodes in a walk is referred to as the {\it duration of the walk}.
 
\begin{figure}[h!]
    \centering
   \includegraphics[scale=0.45]{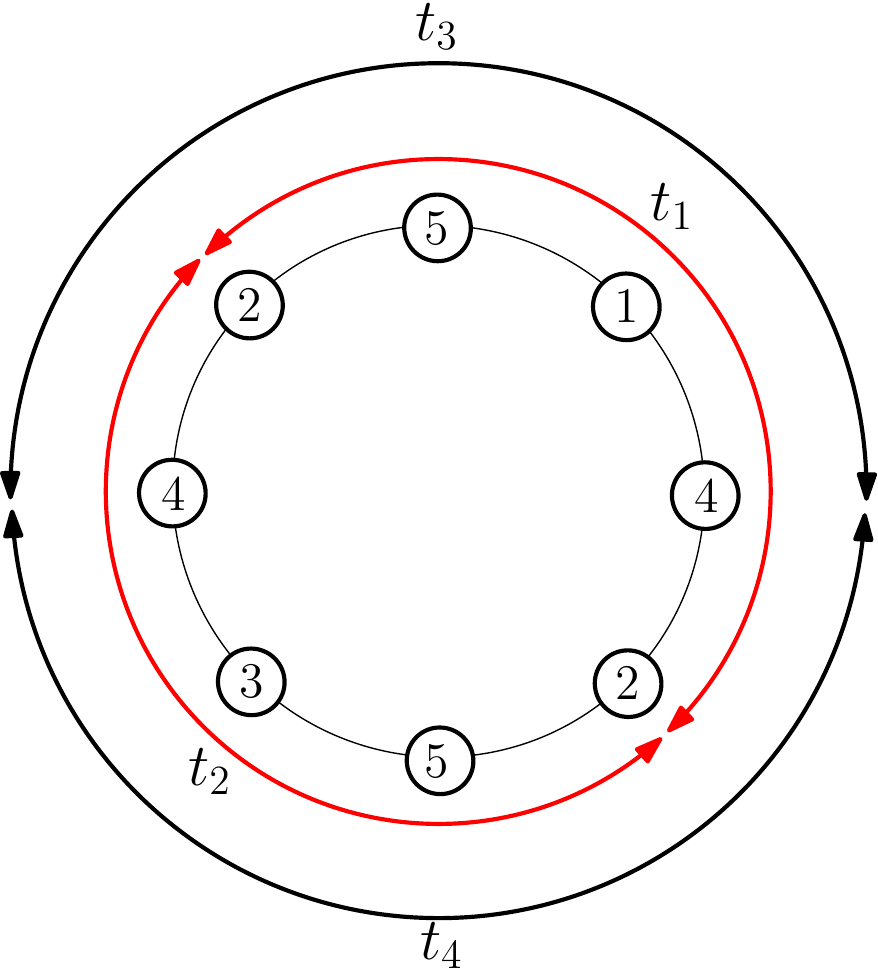}
    \caption{Figure depicting the revisit times of targets 2 and 4 in the walk $\mathcal W(8)$ considered in Figure \ref{fig:walk}}
    \label{fig:rt}
\end{figure}

 Given a walk, $\mathcal W(k)$, of $k$ visits, the revisit time of a target $d \in \mathcal T$, denoted by $RT(d, \mathcal W)$, is the maximum time taken between successive visits to the target, when the walk is repeated. For example, consider the walk shown in Figure \ref{fig:rt}. Target 2 is visited twice in the walk. Time between successive visits to target 2 are $t_1 = c(2,5) + c(5,1) + c(1,4) + c(4,2)$ and $t_2 = c(2,5) + c(5,3) + c(3,4) + c(4,2)$ as shown in the figure. Hence, the revisit time for target 2 is max $\{t_1,t_2\}$. Similarly, the revisit time of target 4 is max $\{t_3,t_4\}$, where $t_3 = c(4,2) + c(2,5) + c(5,1) + c(1,4)$ and $t_4 =  c(4,2) + c(2,5) + c(5,3) + c(3,4)$. The maximum revisit time over all the targets is defined as the {\it revisit time of the walk} or {\it walk revisit time} $\mathcal R(\mathcal W(k))$. Note that {\it if a target is visited exactly once in a walk, it has the maximum revisit time, which is also equal to the duration of the walk}. Therefore, in the example considered above, $\mathcal R(\mathcal W(k))$  = $RT(1, \mathcal W)$ = $RT(3, \mathcal W)$, as targets 1 and 3 are visited exactly once in the walk.
 
 Using the aforementioned notation, the problem is stated as follows:

 \emph{Given $k \ge n$ allowed visits for a UAV, find a closed walk of $k$ visits 
 with the minimum walk revisit time.}\\
 
 A walk with the minimum revisit time is referred to as an optimal walk, and is denoted by $\mathcal W^*(k)$, and the minimum possible revisit time for a given $k$ is denoted by $\mathcal R^*(k)$



\section{Terminology}
This section details the terminology used in the rest of the paper for understanding the structure and the construction of an optimal walk. \\


\noindent
{\it \bf Cyclic Permutation}: \\

\begin{figure}[h!]
    \centering
   \includegraphics[scale=0.45]{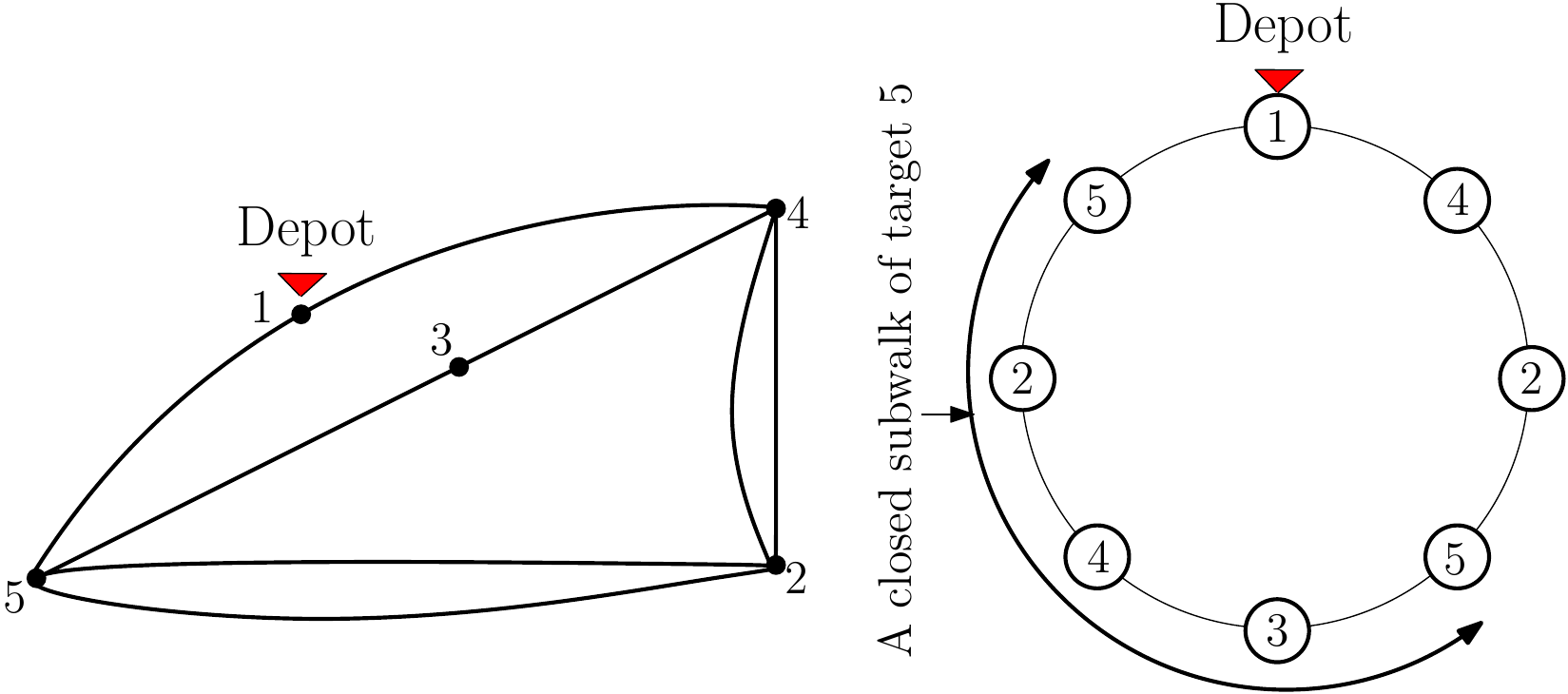}
    \caption{A cyclic permutation ($\bar{\mathcal W} = (1,4,2,5,3,4,2,5,1)$) of the walk $\mathcal W(8)$ depicted in Figures \ref{fig:walk} and \ref{fig:cyclicrep}, with depot at node 1 }
    \label{fig:cyclicwalk}
\end{figure}

A cyclic permutation (or simply, permutation) of a closed walk $\mathcal W$ shifts the first node of the walk to an intermediate node, say $v$, while retaining the order of visits in $\mathcal{W}$; we represent such a cyclic permutation as $\mathcal{C}(\mathcal W, v)$. 
For example, the permutation of $\mathcal W=(v_1,\cdots,v_r,\cdots,v_{k-1},v_1)$ that starts at the intermediate node $v_r$ is defined as $$\mathcal{C}(\mathcal W, v_r):= (v_r,v_{r+1},\cdots,v_{k-1},v_1,v_2,\cdots,v_r).$$ Figure \ref{fig:cyclicwalk} illustrates  $\mathcal{C}(\mathcal W(8), 1)$. \\


\noindent
{\it \bf Concatenation}:\\

\begin{figure}[h!]
\centering

\subfigure[Concatenation of two walks with 5 visits each]{
    \includegraphics[scale=0.4]{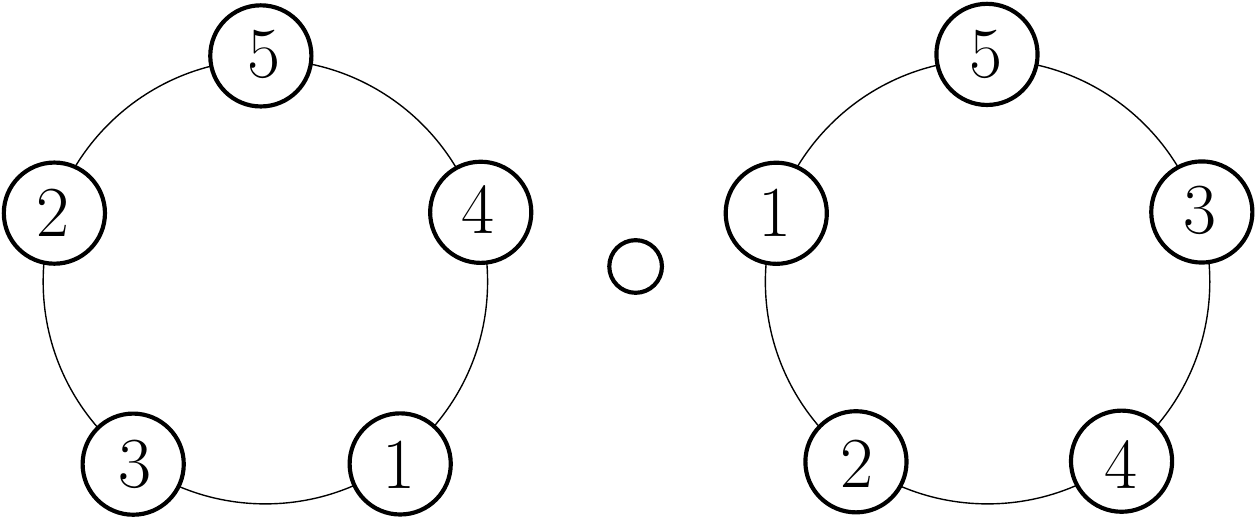}
    \label{fig:concting}
}

\subfigure[A walk with 10 visits obtained by concatenation the walks in \ref{fig:concting}]{
    \includegraphics[scale=0.4]{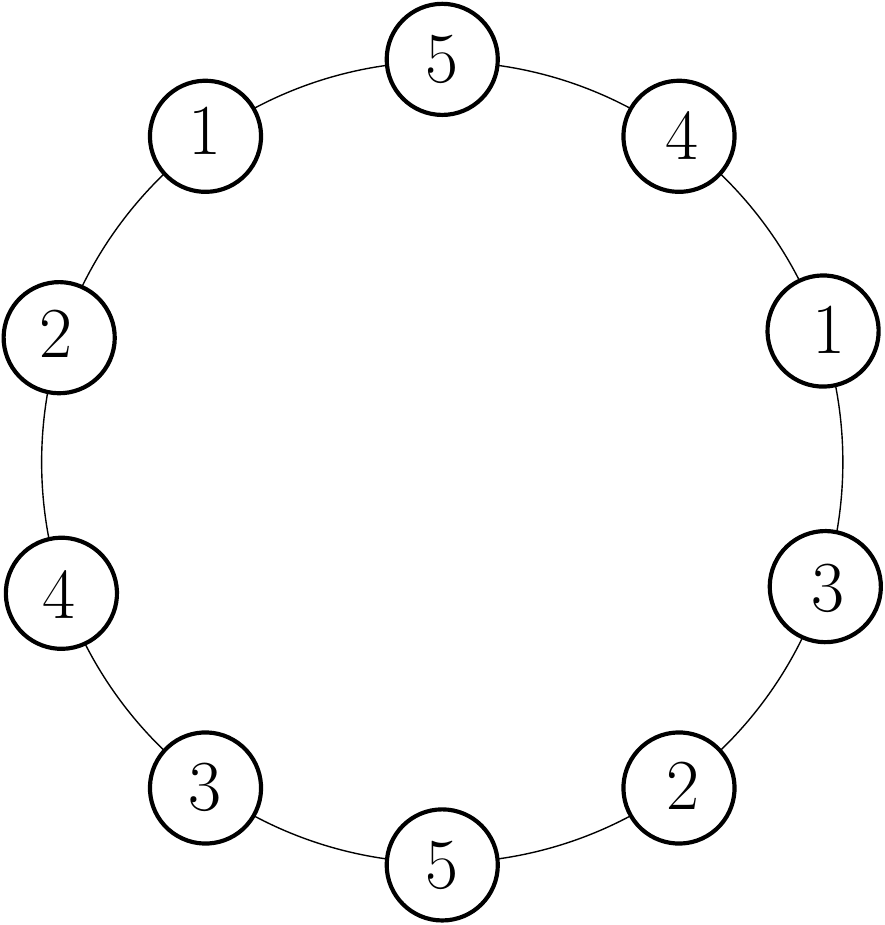}
    \label{fig:concted}
}

\caption{Concatenation of two walks with 5 visits each to form a walk with 10 visits}
\label{fig:def}
\end{figure}

Concatenation of two {\it closed} walks $\mathcal W_1=(d, v_1,\cdots,v_{k},d)$ and $\mathcal W_2=(d,u_1,\cdots,u_{l},d)$ 
is defined as $\mathcal W_1 \circ \mathcal W_2:=(d, v_1,\cdots,v_{k},d,u_1,\cdots,u_{l}, d)$. Figure \ref{fig:concted} shows a walk obtained by concatenating walks in Figure \ref{fig:concting}. \\

It is easy to see that the revisit time of any node does not change when a closed walk is concatenated with itself multiple times. Concatenation of walks is important in constructing feasible walks to the problem corresponding to higher number of visits.\\ 

\begin{figure}[h!]
    \centering
    \includegraphics[scale=0.4]{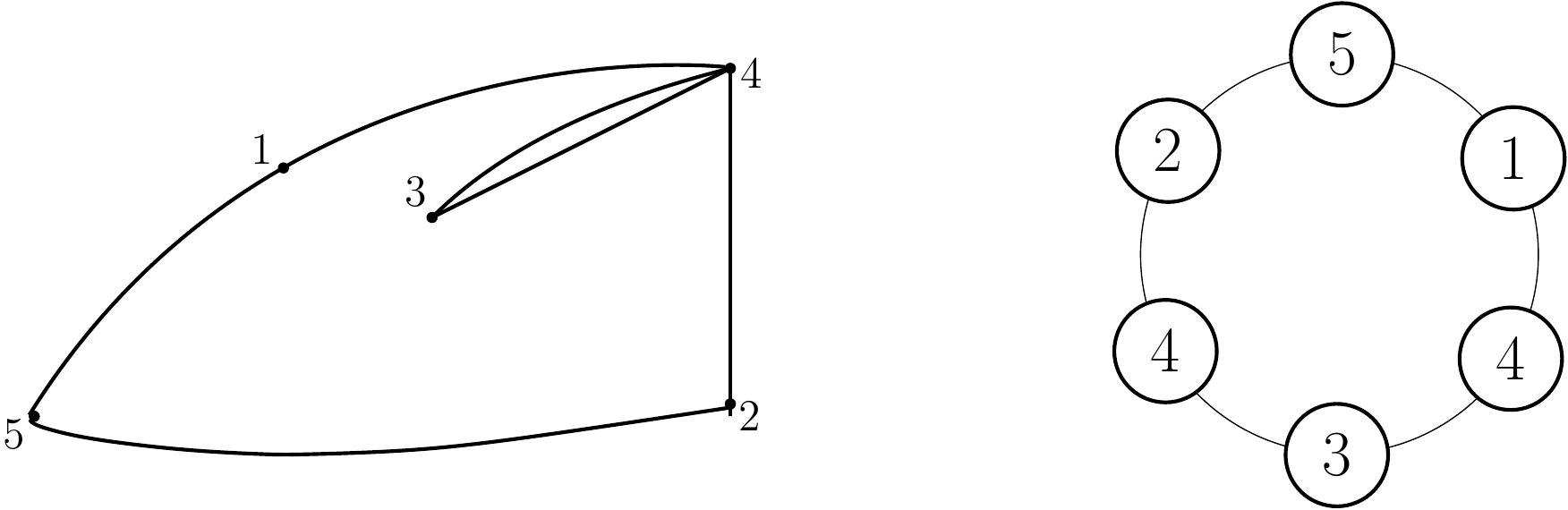}
    \caption{A shortcut walk $\mathcal W(6)$ obtained by shortcutting visits to targets 2 and 5 from the walk $\mathcal W = (5,1,4,2,5,3,4,2,5)$}
    \label{fig:shortcut}
\end{figure}

\noindent
{\it \bf Shortcut Walk}:\\
Given a closed walk $\mathcal W$, removal of a visit to any target $u \in \mathcal W$ is referred to as shortcutting a visit to $u$ from $\mathcal W$. In this article, we refer to a shortcut walk of $\mathcal W$ as a {\it closed walk} obtained by shortcutting visits from $\mathcal W$, but retaining (not shortcutting) the last visits to each target. For example, $\mathcal W(8) = (5,1,4,2,5,3,4,2,5)$ (shown in Figure \ref{fig:walk}) is a closed walk with 8 visits, with two visits to each of targets 2, 4 \& 5, and one visit to each of targets 1 \& 3. A shortcut walk, $\mathcal W(6) = (5,1,4,3,4,2,5)$, of $\mathcal W(8)$ is obtained by shortcutting the first visits to targets 2 \& 5, and retaining the last visits to all the targets in the order they appear in $\mathcal W(8)$, as shown in Figure \ref{fig:shortcut}. 
We remind the readers that shortcutting the last revisits to targets 2, 4 \& 5  or visits to targets 1 \& 3 is not allowed in our definition. In the former case, we cannot shortcut the last revisit to a target; in the latter case, shortcutting does not result in a walk as the shortcut targets are not visited in the walk!


\noindent
{\it \bf Subwalk}: \\

\begin{figure}[h!]
    \centering
    \includegraphics[scale=0.45]{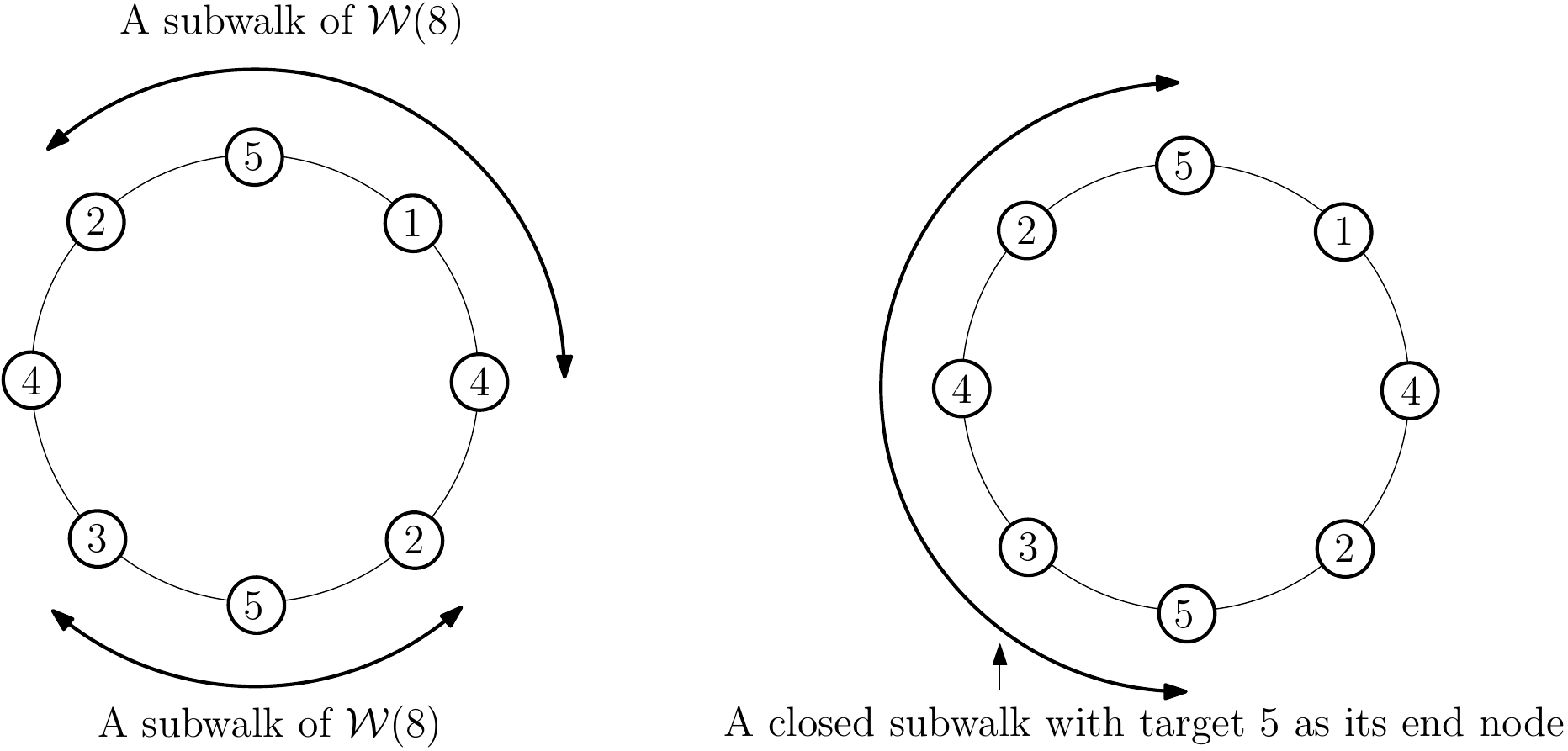}
    \caption{Figure depicting subwalks and closed subwalks of the walk $\mathcal W(8) = (5,1,4,2,5,3,4,2,5)$.}
    \label{fig:subwalk}
\end{figure}

A subwalk $\mathcal W_c$ of a given closed walk $\mathcal W$ is a subsequence obtained by removing certain nodes from the walk, yet retaining the order of visits. A subwalk need not span all the targets and hence, need not be a closed walk. For example, consider the closed walk $\mathcal W(8)$ shown in Figure \ref{fig:subwalk}. The subsequences (2,5,1,4), (2,5,3) are subwalks of $\mathcal W(8)$. A subwalk with the same first and last nodes is referred to as a closed subwalk. Figure \ref{fig:subwalk} shows a closed subwalk of $\mathcal W(8)$ with target 5 as its initial and terminal nodes.

If the terminal node of a subwalk $\mathcal S_1 = (d, v_1, \ldots, v_l)$ is the initial node of a subwalk $\mathcal S_2 = (v_l, v_{l+1}, \ldots, v_q)$, they can be concatenated as $\mathcal S_1 \circ \mathcal S_2 = (v_1, \ldots, v_l, \ldots, v_q)$. Note $\mathcal S_2 \circ \mathcal S_1$ may not be possible unless $v_q = d$. 

The time taken to traverse through all the nodes of a subwalk $\mathcal W_c:=(v_1,\cdots,v_k)$ is referred to as the travel time of the subwalk, and is given by the expression $T(\mathcal W_c) = \sum_{i=1}^{k-1} c(v_i,v_{i+1})$. The travel time of a subwalk is similarly also referred to as the duration of the subwalk.





In some closed subwalks, the end node is not visited in between; we refer to such an end node as a {\it terminus}. For example, $(5,1,4,2,5), (5,3,4,2,5)$ are closed subwalks of $\mathcal{W}(8)$ with node $5$ as the terminus. Given a node $d$ that is visited $l$ times in a closed walk $\mathcal{W}$, it is easy to see that the cyclic permutation $\mathcal C(\mathcal W, d)$ can be decomposed into $l$ closed subwalks with each of them having $d$ as its terminus, i.e., 
 $$ \mathcal C(\mathcal W, d) = \mathcal W_1\circ \mathcal W_2 \cdots \circ \mathcal W_l,$$
 and $d$ is the terminus of closed subwalks $\mathcal W_1, \mathcal W_2, \ldots, \mathcal W_l$. 
Note $T(\mathcal W_1), T(\mathcal W_2), \ldots, T(\mathcal W_l)$ respectively indicate time between successive revisits to target $d$. The revisit time of target $d$ in $\mathcal W$ is 
 $$RT(d,\mathcal W) :=   \max_{1 \le i \le l} \quad T(\mathcal W_i). $$
The revisit time $\mathcal R(\mathcal W)$ of a given walk $\mathcal W$ is the maximum revisit time among all the targets, i.e., $\mathcal R(\mathcal W)=\max_{1 \leq d \leq n} RT(d,\mathcal W)$. Note that {\it the revisit time of a walk is equal to the revisit time of any of the cyclic permutations of the walk}. 

{\it A decomposition of a walk $\mathcal W(k)$ with respect to a vertex $v$} is also useful in analyzing revisit times when concatenating two different walks; suppose $v$ is visited $r$ times in a walk $\mathcal W(k) = (d=v_1, v_2, \ldots, v_{k_1}, \ldots, v_{k_2}, \ldots, v_{k_r}, \ldots, v_1=d)$, i.e., $v_i \ne v$ except for $v_{k_1} = v_{k_2} = \ldots = v_{k_r} = v$; the decomposition of $\mathcal W(k)$ with respect to $v$ is defined as:
\begin{align*}
\mathcal W(k) = \underbrace{(v_1, \ldots, v_{k_1})}_{\mathcal S_1} \circ 
\underbrace{(v_{k_1}, \ldots, v_{k_2})}_{\mathcal W_1}
\circ \underbrace{(v_{k_2}, \ldots, v_{k_3})}_{\mathcal W_2}
\circ  \\ \ldots \circ  \underbrace{(v_{k_{r-1}}, \ldots, v_{k_r})}_{\mathcal W_{r-1}}
\circ \underbrace{(v_{k_r}, \ldots, v_1)}_{\mathcal S_2},
\end{align*}
where $\mathcal W_1, \ldots, \mathcal W_{r-1}$ are subwalks with $v$ as terminus and $\mathcal S_1, \mathcal S_2$ are subwalks with different initial and terminal nodes. Let $\mathcal W = (4,3,5,2,3,1,5,2,3,4,5,1,2,4)$ be a walk of $5$ targets with 13 visits; decomposition of $\mathcal W$ with respect to target $5$ is 
$(4,3,5) \circ (5,2,3,1,5) \circ (5,2,3,4,5) \circ (5,1,2,4)$. 

 Before we proceed to the next section, we introduce the notion of a binding subwalk: Let $\mathcal W(k)$ denote a feasible solution and $\mathcal W^{*}(k)$ represent an optimal solution to the persistent surveillance problem with $k~ (\geq n)$ visits. \\

\noindent{\bf Definition (Binding Subwalk):} A closed subwalk $\mathcal W_b$ is defined to be a binding subwalk if there is a node $d$ such that 
\begin{itemize}
    \item $d$ is the terminus of $\mathcal W_b$, and 
    \item $T(\mathcal W_b) = \mathcal R(\mathcal W^*(k)).$
\end{itemize}

It is easy to see that  $RT(d,\mathcal W^*(k)) = T(\mathcal W_b) = R(\mathcal W^*(k)).$ 

The next section is dedicated to the task of studying the properties of optimal walks.


\section{Properties of optimal walks} \label{sec:proofs}

First, we show that any binding subwalk of an optimal walk must contain visits to all the targets.\\

\begin{lemma}\label{lemma:binding}
{\it  Let $\mathcal W_b$ be a binding subwalk of $\mathcal W^{*}(k)$ for $k\geq n$. Each target is visited at least once in $\mathcal W_b$.}
\end{lemma}

\begin{figure}[h!]
    \centering
    \includegraphics[scale=0.4]{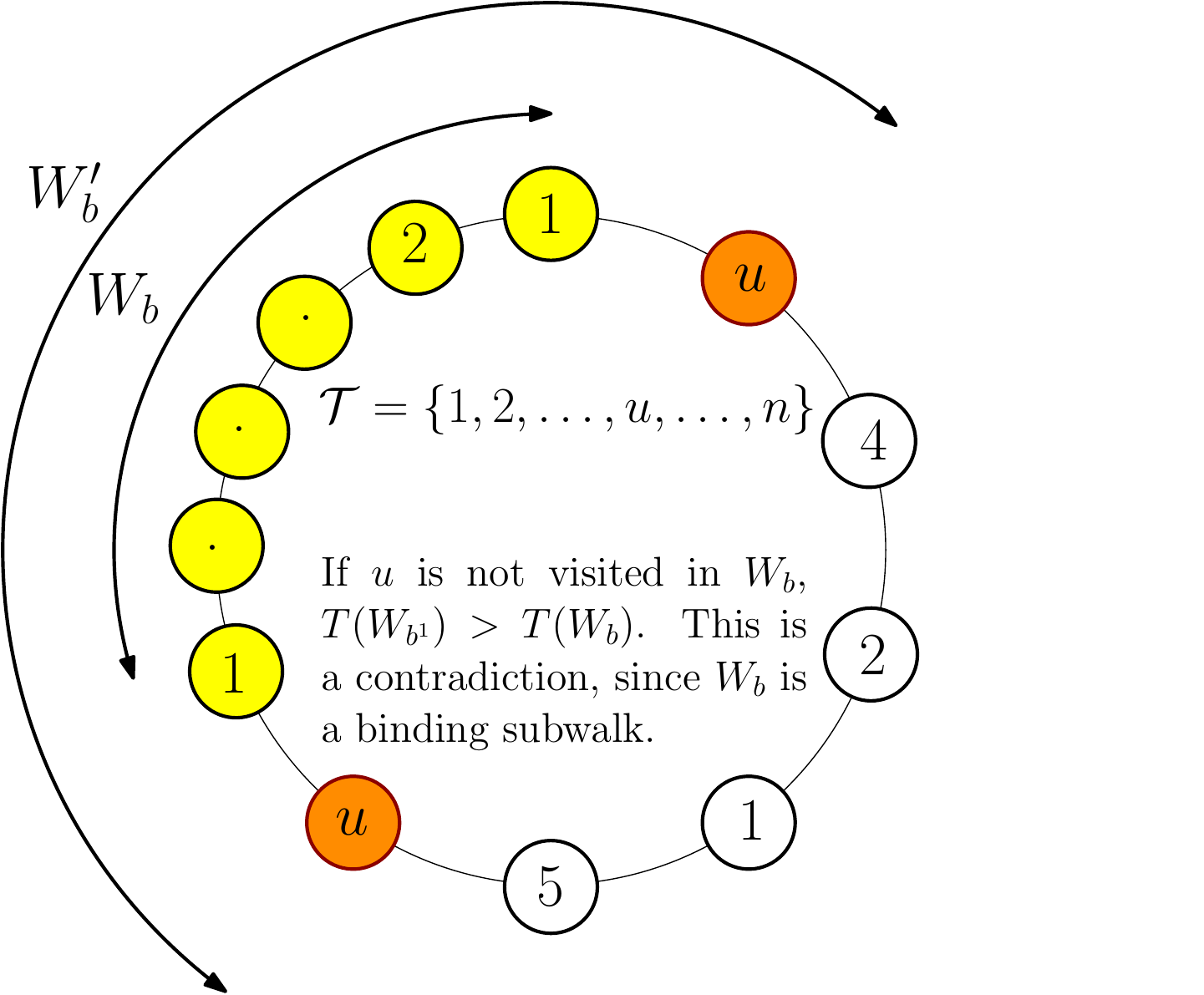}
    \caption{A subwalk $\mathcal W_b$ (shown in yellow) with a missing node $u$ cannot be binding, since there exists another subwalk $\mathcal W'_b$ (with $u$ as its terminus) with a larger travel time}
    \label{fig:lbn}
\end{figure}

\IEEEproof
If a target $u$ is not visited in $\mathcal W_b$, then the revisit time for $u$ must be greater than the time required to traverse $\mathcal W_b$ as shown in Figure \ref{fig:lbn}.
This is because we can find a subwalk $\mathcal W_b'$ with $u$ as its terminus that contains $\mathcal W_b$. However, this contradicts $\mathcal W_b$ being a binding subwalk. Therefore, each target must be visited at least once in $\mathcal W_b$.\\
\endIEEEproof

In essence, $\mathcal W_b$ itself is a walk with atmost $k$ visits.

This result leads to a lower bound on the optimal revisit time. In the following lemma, we prove that the optimal revisit time is lower bounded by the cost of an optimal TSP tour over the targets.\\

\begin{lemma}\label{lemma:lowerbound}
{\it  $\mathcal R(\mathcal W^{*}(k)) \geq \mathcal R(\mathcal W^{*}(n)) = TSP^*$ where $TSP^*$ denotes the cost of an optimal TSP tour visiting all the $n$ targets.}
\end{lemma}

\IEEEproof
Any feasible solution to the TSP is also a feasible solution to the persistent surveillance problem with $n$ visits where each target is visited exactly once, and vice versa. Therefore, $\mathcal R(\mathcal W^{*}(n)) = TSP^*$.

    From Lemma \ref{lemma:binding}, for $k>n$, each target is visited at least once in a binding subwalk $\mathcal W_b$. Shortcut any target that is visited more than once in $\mathcal W_b$ to obtain a tour, $\mathcal W_{tour}$. As the travel times satisfy the triangle inequality, short cutting any visit will not increase the revisit time, $i.e.$, $T(\mathcal W_b) \geq \mathcal R(\mathcal W_{tour})$. Therefore, $\mathcal R(\mathcal W^{*}(k)) = T(\mathcal W_b) \geq \mathcal R(\mathcal W_{tour}) \geq TSP^*$. \\
\endIEEEproof


 From Lemma \ref{lemma:lowerbound}, $TSP^*$ is a lower bound for every walk revisit time. In some cases, this bound can be tightened, especially when the number of visits, $k$, is not an integral multiple of $n$. \\

\begin{lemma} \label{lemma:lbgeneral}
Consider a walk $\mathcal W(k)$ with $k \geq n$ visits. One can express $k$ in the quotient-remainder form as $pn+q$, where $p,q \in Z_+$, $p \geq 1$, and $0 \leq q \leq n-1$. Then, $\mathcal W(k)$ contains a closed subwalk $\mathcal W_s$ with a terminus, with at least $n+\myceil{\frac{q}{p}}$ visits. Consequently, $\mathcal R^*(k) \geq \mathcal R^*(n+\myceil{\frac{q}{p}})$.\\
\end{lemma}

\IEEEproof
Suppose the maximum number of visits in a closed subwalk with a terminus is at most $n+\myceil{\frac{q}{p}} - 1$. Then, we show that each target must be visited at least $p+1$ times in the walk as follows: In any given walk, the maximum number of visits between consecutive revisits to a target is lower bounded by the average number of visits between consecutive revisits to the same target. If a target $v$ is visited at most $p$ times, the maximum number of visits between consecutive revisits to $v$ is at least $\frac{pn+q}{p} (> n+ \myceil{\frac{q}{p}} - 1$). However, if each target is visited at least $p+1$ times, the total number of visits in the walk, $k$, is at least $(p+1)n > pn+q$, which contradicts the hypothesis. \\

Hence, there exists a closed subwalk $\mathcal W_s$ with a terminus, that has at least $n+\myceil{\frac{q}{p}}$ visits. Without loss of generality, one can assume that $\mathcal W_s$ spans all the targets. (If not, 
one can find a target $u \notin \mathcal W_s$, such that it is the terminus of another closed subwalk that contains $\mathcal W_s$ and spans all the targets
). By definition, $\mathcal R(\mathcal W(k)) \geq T(\mathcal W_s)$, and from triangle inequality, it follows that $T(\mathcal W_s) \geq \mathcal R^*(n+\myceil{\frac{q}{p}})$. Therefore, $\mathcal R(\mathcal W(k)) \geq T(\mathcal W_s)$.  Since this is true for any walk with $k$ visits, we have $\mathcal R^*(k) = \mathcal R(\mathcal W^*(k)) \geq \mathcal R^*(n+\myceil{\frac{q}{p}})$.

\endIEEEproof

\noindent{\bf Remark:} If $p \ge n-1$, it is clear that $q \ne 0 \Rightarrow \myceil{\frac{q}{p}} = 1$ and consequently, if $k$ is not an integral multiple of $n$ and $p \ge n-1$, we have $\mathcal{R}^*(k) \ge \mathcal{R}^*(n+1)$. We will use this fact in Theorem 2.\\

In the subsequent sections, we will construct feasible walks by concatenating smaller walks in such a way that the maximum revisit time equals the lower bound found in this section, thereby demonstrating optimality. The key result in the next section is that concatenating a binding subwalk with some of its shortcut walks and subsequently replacing the binding walk with  the concatenated walk is not going to increase the maximum revisit time. This result is important in that walks with higher number of visits can be constructed without increasing the maximum revisit time. 

\subsection{Properties of Concatenating Feasible Solutions}



\begin{figure}[h!]
\centering

\subfigure[]{
    \includegraphics[scale=0.32]{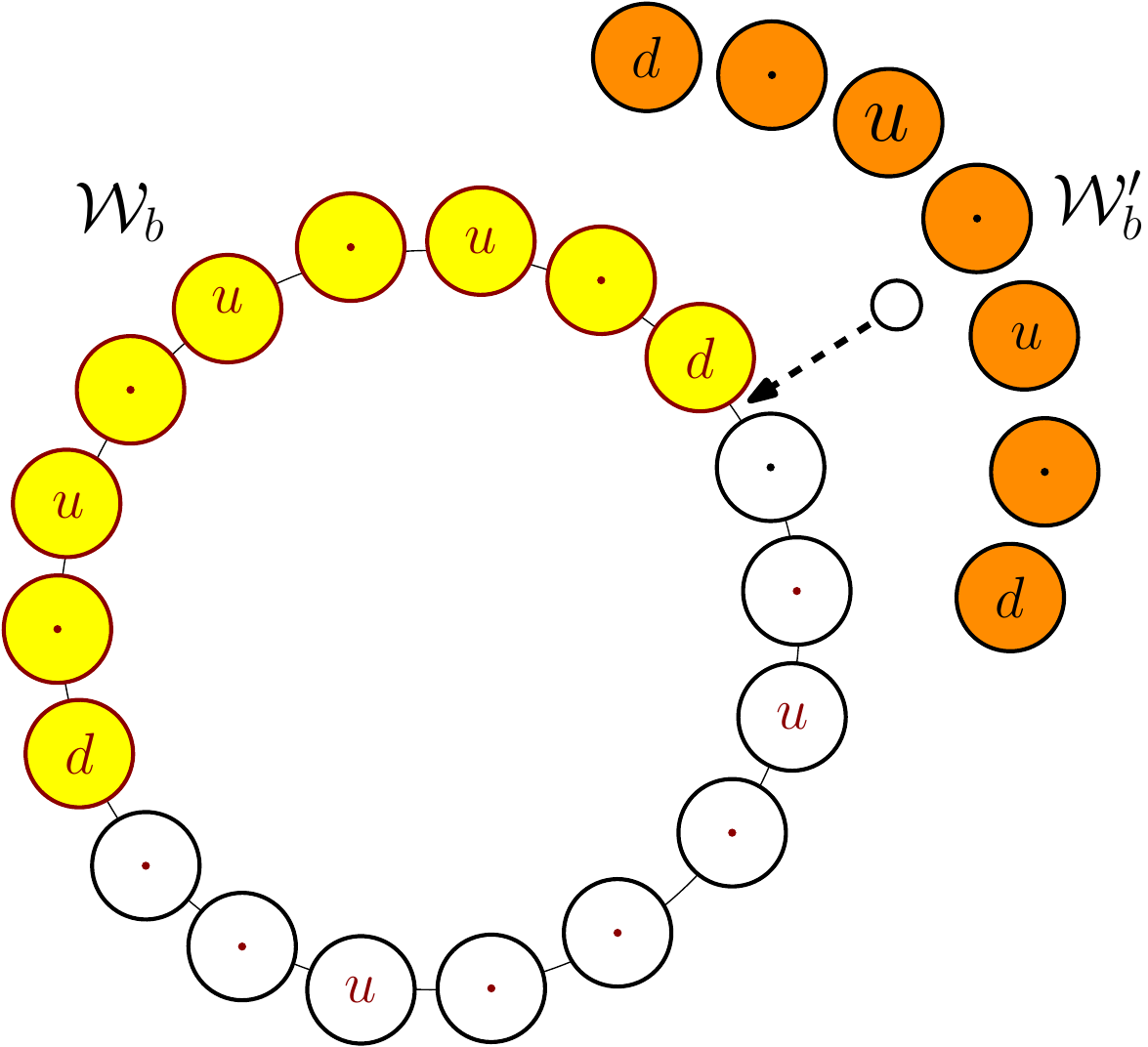}
    \label{fig:concatgen1}
}
\subfigure[]{
    \includegraphics[scale=0.25]{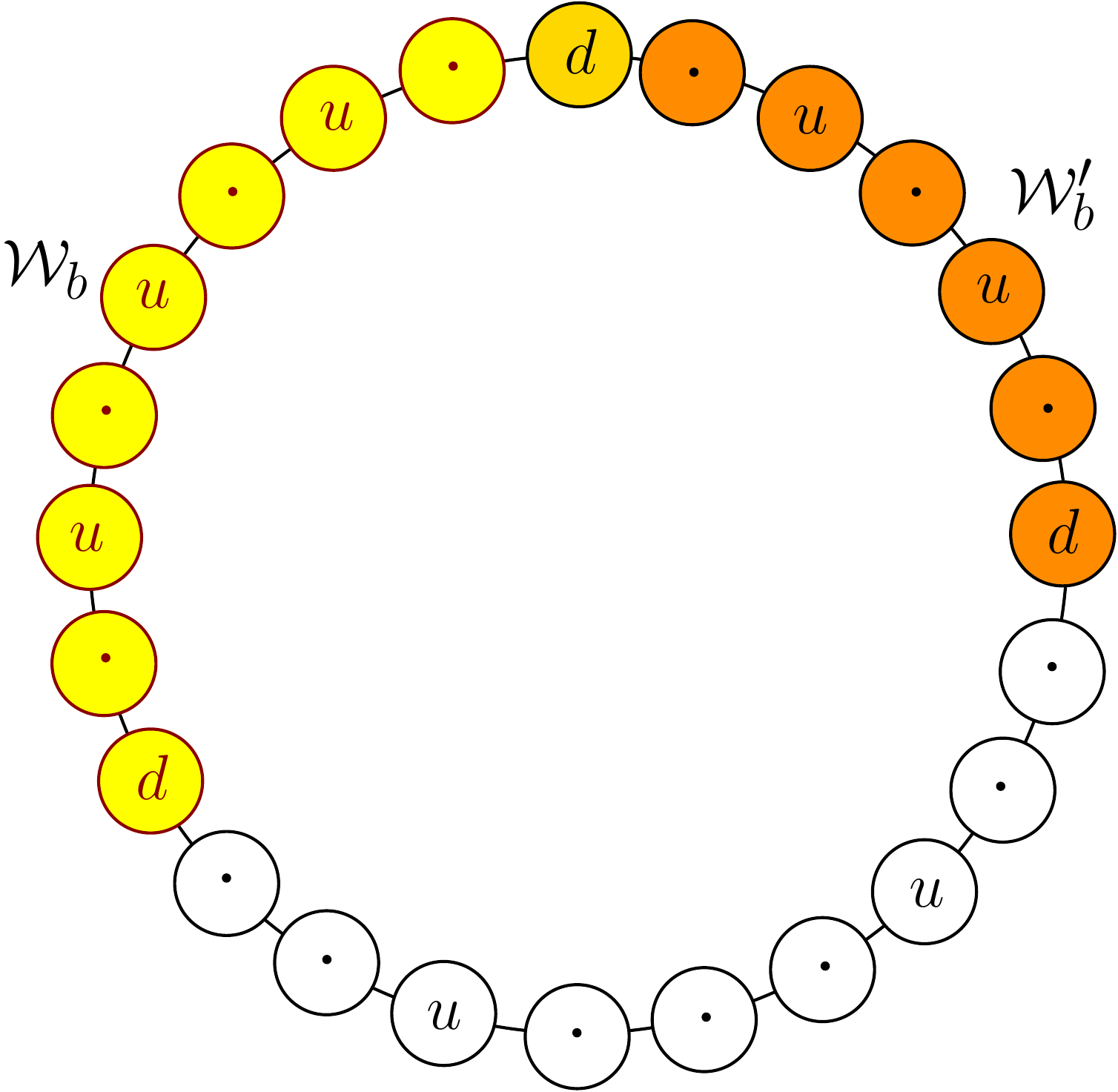}
    \label{fig:concatgen2}
}

\caption[]{A closed walk $\mathcal W$ is shown on the left, with its binding subwalk $\mathcal W_b$ colored yellow. A subwalk $\mathcal W'_b$ (colored orange) is constructed from $\mathcal W_b$, by shortcutting visits to target $u$ (retaining the last visit to $u$). Figure on the right shows the concatenated walk $\bar{\mathcal W}$, as discussed in Theorem \ref{theorem:concatgeneral}.}

\label{fig:concatgen11}
\end{figure}

The procedure to extend a walk into a larger feasible solution is given by the following lemma, and is depicted in Figure \ref{fig:concatgen11}.\\

\begin{theorem} \label{theorem:concatgeneral}
{\it Let $\mathcal W_b$ be a binding subwalk and $S_1$, $S_2$ be subwalks of $\mathcal W$ such that $\mathcal W = S_1 \circ \mathcal W_b \circ S_2$. Let $\mathcal W'_b$ be a closed subwalk obtained by shortcutting visits from $\mathcal W_b$, but retaining the last visits to targets. Let a closed walk $\bar{\mathcal W}$ be formed by concatenating $\mathcal W$ with $\mathcal W'_b$ as follows: $\bar{\mathcal W} =S_1 \circ \mathcal W_b \circ \mathcal W'_b \circ S_2$. Then, $\mathcal R(\bar{\mathcal W}) = \mathcal R(\mathcal W)$.\\ }
\end{theorem}

\IEEEproof See Appendix \ref{sec:addproofs}.\\


Next, we present two lemmas involving the cyclic permutation and concatenation of walks. The first of the two deals with the revisit time of a closed walk concatenated with itself an arbitrary number of times.\\

\begin{lemma}\label{lemma:sameconcat}
{\it Let a closed walk $\bar{\mathcal W}$ be formed by concatenating $\mathcal W(k)$ with itself $p$ times as follows:
     $$\bar{\mathcal W} :=  \underbrace{\mathcal W(k)\circ \cdots \circ \mathcal W(k)}_{p ~times}. $$
Then, for any node $v_r$
\begin{itemize}
    \item 
    $\mathcal{C}(\bar{\mathcal{W}},v_r) = \underbrace{\mathcal{C}(\mathcal{W}(k),v_r) \circ 
    \cdots \circ \mathcal{C}(\mathcal{W}(k),v_r)}_{p \; times}.$
    
    \item $RT(v_r,\bar{\mathcal W}) = RT(v_r,\mathcal W(k))$, 
\end{itemize}
and thus ${\mathcal R}(\bar{\mathcal W}) = \mathcal R (\mathcal W(k)).$\\
  }
\end{lemma}
\IEEEproof See Appendix \ref{sec:addproofs}.\\
\endIEEEproof

{The next lemma involves concatenating a walk with its shortcut walk and is crucial for piecing together smaller feasible walks to build a bigger walk while not increasing the maximum walk revisit time.\\}

\begin{lemma}\label{lemma:permut}
{\it Let $\mathcal W(k)$ be a closed walk with $n+1 \le k \le 2n-1$ visits, and $d$ be a target that is visited exactly once in $\mathcal W(k)$. Let $\mathcal W(k-1)$ be a walk formed by shortcutting a revisit from  $\mathcal W(k)$. Consider a closed walk $\bar{\mathcal W}$ formed by concatenating $\mathcal W(k)$ for $q~ (\geq 1)$ times and $\mathcal W(k-1)$ for $p$ times as follows:
      $$\bar{\mathcal W} := \underbrace{\mathcal W(k)\circ ..\circ \mathcal W(k)}_{q ~times}\circ \underbrace{\mathcal W(k-1)\circ .. \circ \mathcal W(k-1)}_{p ~times}, $$
Then, 
\begin{multline*}
\mathcal{C}(\bar{\mathcal{W}},d) = 
\underbrace{\mathcal{C}(\mathcal{W}(k),d) \circ \cdots \circ \mathcal{C}(\mathcal{W}(k),d)}_{q \; times} \circ \\ \underbrace{\mathcal{C}(\mathcal{W}(k-1),d) \circ \cdots \circ \mathcal{C}(\mathcal{W}(k-1),d)}_{p \; times}. 
\end{multline*}
  }
\end{lemma}
\IEEEproof
See Appendix \ref{sec:addproofs}\\
\endIEEEproof
\noindent{\bf Remark:} This lemma can be used to construct a feasible walk with $K$ visits as follows: Let $K= rn+s$ for some integers $r>1, s\ge0$ and $n$ being the number of targets. If $K \ge n(n-1)$, then $K = r n + s$, where $r \ge n-1, s \in \{0, 1, 2, \ldots, n-1\}$. Set $k = n+1$, $p=s$ and $q= r-s$ so that $pk+q(k-1)= p(n+1)+q(n) = (p+q)n + p = rn + s = K$; using the above lemma, one can construct a feasible walk for $K$ visits from the feasible walk of $n+1$ visits and its shortcut walk of $n$. The following lemma shows that building such a bigger walk can be done without increasing the maximum walk revisit time. 

\begin{lemma} \label{lemma:concatv} 
{\it Let $\mathcal W^*(k)$ be an optimal solution with $n+1 \le k \le 2n-1$ visits, and $\mathcal W(k-1)$ be a shortcut walk of $\mathcal W^*(k)$. 
Let a closed walk $\bar{\mathcal W}$ be formed by concatenating $\mathcal W^*(k)$ for $q~ (\geq 1)$ times and $\mathcal W(k-1)$ for $p$ times as follows:
     $$\bar{\mathcal W} := \underbrace{\mathcal W^*(k)\circ ..\circ \mathcal W^*(k)}_{q ~times}\circ \underbrace{\mathcal W(k-1)\circ .. \circ \mathcal W(k-1)}_{p ~times}, $$
Then the maximum revisit time for $\bar{\mathcal W}$ is equal to $\mathcal R(\mathcal W^*(k))$ i.e., \[\mathcal R(\bar{\mathcal W}) = \mathcal R(\mathcal W^*(k)).\]  }
\end{lemma}





\IEEEproof
 Since $k \le 2n-1$, there is at least one node, $d$, in $\mathcal W^*(k)$ that is visited exactly once. Consider a cyclic permutation $\mathcal C(\mathcal W^*(k),d)$ of $\mathcal W^*(k)$. Let a closed walk $\mathcal W_q$ be formed by concatenating $\mathcal C(\mathcal W^*(k),d)$ for $q$ times as follows:\\
$$\mathcal W_q = \underbrace{\mathcal C(\mathcal W^*(k),d)\circ ..\circ \mathcal C(\mathcal W^*(k),d)}_{q ~times}.$$ 
From Lemma \ref{lemma:sameconcat} it follows that $\mathcal R(\mathcal W_q) = \mathcal R(\mathcal C(\mathcal W^*(k),d)) = \mathcal R(\mathcal W^*(k))$. As $d$ is visited exactly once in $\mathcal W^*(k)$, we have $T(\mathcal C(\mathcal W^*(k),d)) = \mathcal R(\mathcal C(\mathcal W^*(k),d))$. Moreover, $d$ is the terminus of $\mathcal C(\mathcal W^*(k),d)$. Therefore, $\mathcal C(\mathcal W^*(k),d)$ is a binding subwalk of itself. Because $ T(\mathcal C(\mathcal W^*(k),d)) = \mathcal R(\mathcal C(\mathcal W^*(k),d)) = \mathcal R(\mathcal W_q)$, $\mathcal C(\mathcal W^*(k),d)$ is also a binding subwalk of $\mathcal W_q$. 

Now consider $\mathcal W(k-1)$, which is formed by shortcutting a revisit from $\mathcal W^*(k)$. Note that $d$ is visited exactly once in $\mathcal W(k-1)$. $\mathcal{C}(\mathcal W(k-1),d)$, is a shortcut walk of $\mathcal{C}(\mathcal W^*(k),d)$, which is a binding subwalk of $\mathcal W_q$. It follows from Theorem \ref{theorem:concatgeneral} that the revisit time of a walk formed by concatenating $\mathcal W_q$ with $\mathcal{C}(\mathcal W(k-1),d)$, is $\mathcal R(\mathcal W^*(k))$. Since $\mathcal{R}(\mathcal W_q \circ \mathcal{C}(\mathcal W(k-1),d)) = \mathcal R(\mathcal W^*(k))$, $\mathcal{C}(\mathcal W(k),d)$ is also a binding subwalk of $\mathcal W_q \circ \mathcal{C}(\mathcal W(k-1),d)$. Therefore, concatenating $\mathcal W_q \circ \mathcal{C}(\mathcal W(k-1),d)$ with $\mathcal{C}(\mathcal W(k-1),d)$ does not change the revisit time, and the process can be repeated for any number of times (say $p$). Hence, if $\mathcal W_p$ is defined as,\\
$$\mathcal W_p = \underbrace{\mathcal C(\mathcal W(k-1),d)\circ ..\circ \mathcal C(\mathcal W(k-1),d)}_{p ~times},$$ 
we have $\mathcal R(\mathcal W_q \circ \mathcal W_p) = \mathcal R(\mathcal W^*(k)).$ From Lemma \ref{lemma:permut}, $ \mathcal W_q \circ \mathcal W_p$ is a cyclic permutation of $\bar{\mathcal W}$, and thus $\mathcal R(\bar{\mathcal W}) = \mathcal R(\mathcal W_q \circ \mathcal W_p) = \mathcal R(\mathcal W^*(k)).$

 \endIEEEproof

With this, we have enough tools to analyze the properties of optimal revisit time. To study the properties of optimal solutions, we split the number of visits into 3 categories: $n \leq k \leq 2n-1$, $2n \le k < n^2-n$ and $k \geq n^2-n$. We begin with the case, $k \geq n^2-n$, and prove that the optimal revisit time is a periodic function of $k$, with a period $n$.


\subsection{Bi-modal Property of  Optimal Revisit Time}
The focus of this subsection is to prove the main result of this article, that after a finite number of visits ($k \geq n^2-n$), the optimal revisit is bi-modal with a period $n$. The two values it takes are: $TSP^*$ when $k$ is an integral multiple of $n$, and $\mathcal R(\mathcal W^{*}(n+1))$ when $k$ is not an integral multiple of $n$.\\



\begin{lemma}\label{lemma:even}
{\it $\mathcal R(\mathcal W^{*}(k)) = TSP^*$ for $k=pn$ where $p$ is any positive integer. }
\end{lemma}
\IEEEproof
Consider the walk $\mathcal W:=\underbrace{\mathcal W^*(n)\circ \cdots \circ \mathcal W^*(n)}_{p ~times} $. $\mathcal W$ is feasible to the problem with $pn$ visits. Using Lemma \ref{lemma:sameconcat},  $\mathcal R(\mathcal W)=\mathcal R(\mathcal W^*(n)) = TSP^*$. Therefore, from Lemma \ref{lemma:lowerbound}, it follows that $\mathcal W$ is also an optimal solution to the problem with $pn$ visits. \\
\endIEEEproof

\begin{lemma}\label{lemma:ubodd}
{\it $\mathcal R(\mathcal W^{*}(k)) \leq \mathcal R(\mathcal W^{*}(n+1))$ for $k > n^2-n$ and $k$ is not an integral multiple of $n$. }
\end{lemma}

\IEEEproof
Consider the optimal walk $\mathcal W^*(n+1)$ with $n+1$ visits. Since exactly one target is visited twice in $\mathcal W^*(n+1)$, shortcut one of the visits to this target to get a new walk $\mathcal W$. Let $k = np+q$, $p\ge n-1$ and $1\le q\le n-1$. Consider the following feasible solution to the problem with $k$ visits:
$$\bar{\mathcal W} = \underbrace{\mathcal W^*(n+1)\circ \cdots \circ \mathcal W^*(n+1)}_{q \; times}\circ \underbrace{\mathcal W \circ \cdots \circ \mathcal W }_{(p-q) \; times}.$$ Note that the total number of visits in $\bar{\mathcal W}$ is $q(n+1) + (p-q)n = np+q = k$. Using Lemma \eqref{lemma:concatv}, we obtain $\mathcal R(\bar{\mathcal W}) = \mathcal \mathcal R(\mathcal \mathcal W^*(n+1))$. Therefore, $\mathcal R(\mathcal W^{*}(k)) \leq \mathcal R(\bar{\mathcal W}) \leq \mathcal R(\mathcal W^{*}(n+1))$.\\
\endIEEEproof

\begin{theorem} \label{theorem:optimal}
{\it For $k\geq n^2-n$, the optimal revisit time for the persistent surveillance problem can be only one of two values, $i.e.$,
$$
\mathcal R(\mathcal W^{*}(k)) =
\begin{cases}
\mathcal R(\mathcal W^{*}(n)),  k=pn, p\textrm{ is an integer}, \\
\mathcal R(\mathcal W^{*}(n+1)), \textrm{ otherwise}.
\end{cases}
$$
}
\end{theorem}

\IEEEproof
The theorem follows from Lemmas \eqref{lemma:even} and \eqref{lemma:ubodd}, and the remark following Lemma \eqref{lemma:lbgeneral}.
\endIEEEproof

\subsection{Monotonicity of Optimal Revisit Time}
In this subsection, we show that the optimal revisit time is a monotonic function of the number of visits, for $n \leq k \leq 2n-1$ 
\begin{lemma}\label{lemma:monotonic}
{\it $\mathcal R(\mathcal W^{*}(k)) \leq \mathcal R(\mathcal W^{*}(k+1))$ for $n \leq k \leq 2n-2$. }
\end{lemma}
\IEEEproof
Consider the optimal solution $\mathcal W^{*}(k+1)$. As $k+1\leq 2n-1$, there is at least one node in $\mathcal W^{*}(k+1)$ that is visited exactly once. Let $d$ denote one of these nodes. Since $d$ is visited exactly once, $RT(d,\mathcal W^{*}(k+1)) = \mathcal R(\mathcal W^{*}(k+1))$. Also, as $k+1>n$, there is at least one node (say $u$) that is visited more than once in $\mathcal W^{*}(k+1)$. Shortcut one of the revisits to $u$ from $\mathcal W^{*}(k+1)$ to obtain a feasible walk $\mathcal W$ to the problem with $k$ visits. As the travel times satisfy the triangle inequality, $RT(d,\mathcal W)\leq RT(d,\mathcal W^{*}(k+1))$. Therefore, $\mathcal R(\mathcal W^{*}(k)) \leq \mathcal R(\mathcal W)=RT(d,\mathcal W) \leq RT(d,\mathcal W^{*}(k+1)) \leq \mathcal R(\mathcal W^{*}(k+1))$.
\endIEEEproof


\subsection{Bounds on Optimal Revisit Time}
In this subsection, we provide a procedure to construct feasible walks for the case $2n < k < n^2-n$, using the optimal solutions for $n < k \leq 2n-1$. The revisit times of these walks match the lower bounds presented earlier in this section, proving optimality of the constructed solutions.\\


\begin{lemma}\label{lemma:ubint} Let $k > n$ be not an integral multiple of $n$, i.e., $k= pn + q$ for some integers $p \geq 1$, $1 \leq q \leq n-1$. Then
{\it $\mathcal R(\mathcal W^{*}(k)) \leq \mathcal R(\mathcal W^{*}(n+\myceil{\frac{q}{p}}))$.}
\end{lemma}

\IEEEproof
It is sufficient to construct a feasible solution $\bar{\mathcal W}$ with $k$ visits, such that $\mathcal R(\bar{\mathcal W}) = \mathcal R(\mathcal W^{*}(n+\myceil{\frac{q}{p}}))$. Consider an optimal walk $\mathcal W^*(n+\myceil{\frac{q}{p}})$ with $n+\myceil{\frac{q}{p}}$ visits, and a corresponding walk $\mathcal W(n+\floor{\frac{q}{p}})$ with $n+\floor{\frac{q}{p}}$ visits as follows:
\begin{enumerate}
\item $\mathcal W(n+\floor{\frac{q}{p}})$ is obtained by shortcutting a visit to a repeated target if $q \ne p$.
\end{enumerate}

Let $q=ps+r$ for some integers $s \ge 1$ and $0 \leq p-1$.
For the sake of convenience, let us refer to $\mathcal W^*(n+\myceil{\frac{q}{p}})$ as $\mathcal W^*$, and $\mathcal W(n+\floor{\frac{q}{p}})$ as $\mathcal W$. Then, a feasible solution to the problem can be constructed by concatenating $\mathcal W^*(n+\myceil{\frac{q}{p}})$ times and $\mathcal W(n+\floor{\frac{q}{p}})$ times as follows:

$$\bar{\mathcal W} := \underbrace{\mathcal W^*\circ \cdots \circ \mathcal W^*}_{r \; times}\circ \underbrace{\mathcal W \circ \cdots \circ \mathcal W}_{(p-r) \; times}.$$

Note that the total number of visits in $\bar{\mathcal W}$ is $r(n+\myceil{\frac{q}{p}}) + (p-r)(n+\floor{\frac{q}{p}}) = np+q = k$; the equality can be inferred from the following cases:
\begin{enumerate}
\item If $q=p$, $\myceil{\frac{q}{p}} = \floor{\frac{q}{p}} = 1$. Hence, $r(n+\myceil{\frac{q}{p}}) + (p-r)(n+\floor{\frac{q}{p}}) = pn+p = pn +q$. 
\item If $q<p$, we have $r = q$, $\myceil{\frac{q}{p}} = 1$ and $\floor{\frac{q}{p}} = 0$. Therefore, $r(n+\myceil{\frac{q}{p}}) + (p-r)(n+\floor{\frac{q}{p}}) = q(n+1) + (p-q)n = pn+q$.
\item If $q>p$, we have $\myceil{\frac{q}{p}} = s+1$ and $\floor{\frac{q}{p}} = s$. Hence, $r(n+\myceil{\frac{q}{p}}) + (p-r)(n+\floor{\frac{q}{p}}) = r(n+s+1) + (p-r)(n+s) = pn+ps+r = pn+q$.
\end{enumerate}
Since $n+ \myceil{\frac{q}{p}} \leq 2n-1$, from Lemmas \ref{lemma:sameconcat} and \ref{lemma:concatv}, we can conclude $\mathcal R(\bar{\mathcal W}) = \mathcal R(\mathcal W^{*}(n+\myceil{\frac{q}{p}}))$.\\
\endIEEEproof

\begin{theorem} \label{theorem:transient}
For any given number of visits $k \geq n$, $\mathcal R^*(k) = \mathcal R^*(n+\myceil{\frac{q}{p}})$, where $k = pn+q$, $p \geq 1$, $0 \leq q \leq n-1$ and $p, q \in Z_+$.
\end{theorem}

\IEEEproof
The theorem follows from Lemmas \ref{lemma:ubint}, \ref{lemma:lbgeneral} and \ref{lemma:even}.
\endIEEEproof

A consequence of this result is that one needs to solve at most $n$ distinct problems to construct optimal walks for any given $k \geq n$.

We conclude this section with another general result that holds for any $k \geq n$. Given an optimal walk $\mathcal W^*(k)$ with $k$ visits, this result enables one to construct feasible solutions of $k+n$ visits, without increasing the revisit time. This translates to the fact that the optimal revisit time for a problem with $k+n$ visits is upper bounded by that for a problem with $k$ visits. Hence, for every $n$ visits, the optimal revisit time either reduces or stays the same, but does not increase.\\

\begin{corollary}[of Theorem \ref{theorem:concatgeneral}] \label{corollary:vpn}
{\it The optimal revisit time for the persistent surveillance problem with $k$ visits is an upper bound to the one with $k+n$ visits. That is, $\mathcal R(\mathcal W^*(k+n)) \leq \mathcal R(\mathcal W^*(k))$, $\forall k \geq n$.}
\end{corollary}

\begin{figure}[h!]
\centering
\subfigure[]{
    \includegraphics[scale=0.32]{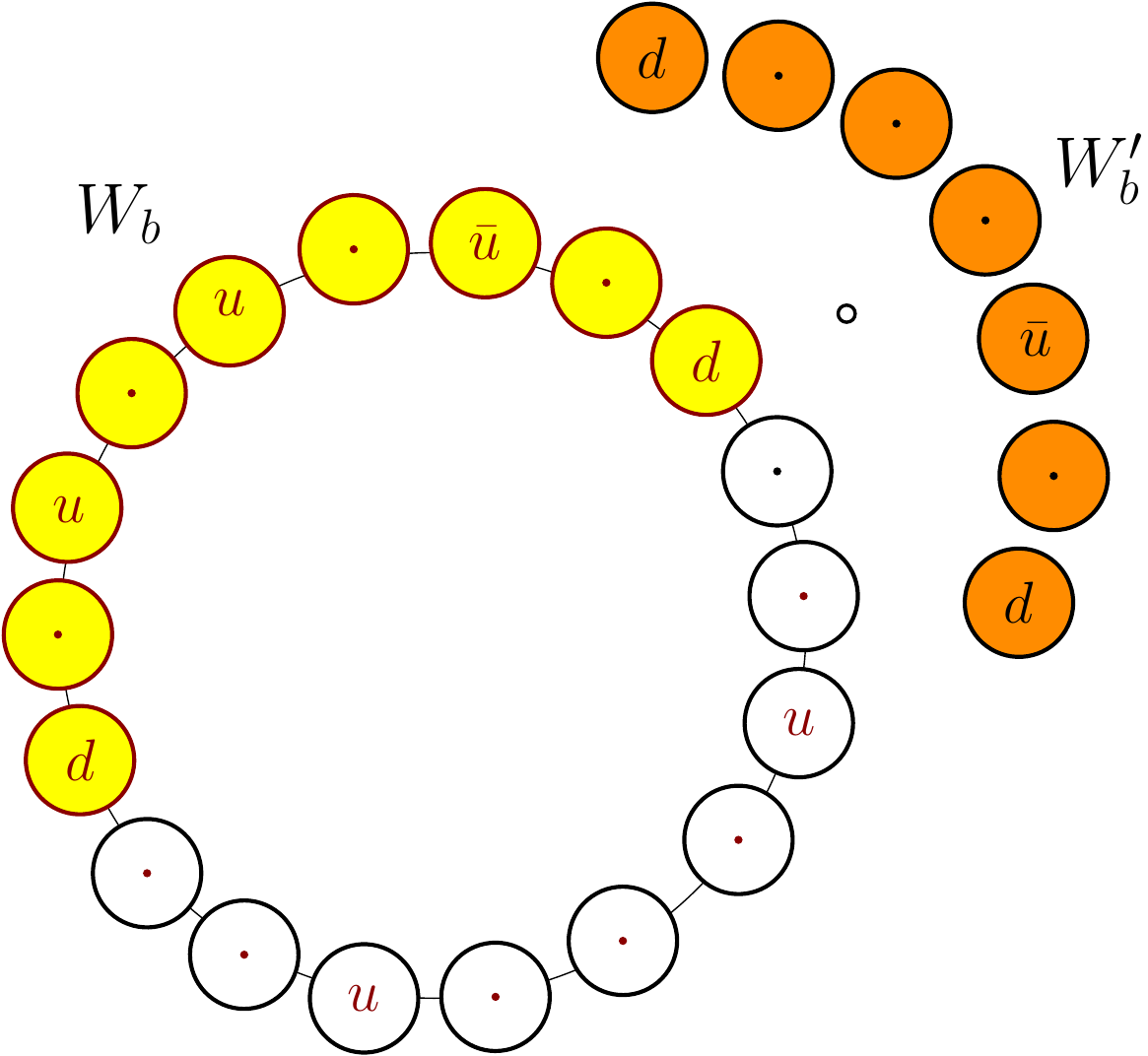}
    \label{fig:vvpn1}
}
\subfigure[]{
    \includegraphics[scale=0.25]{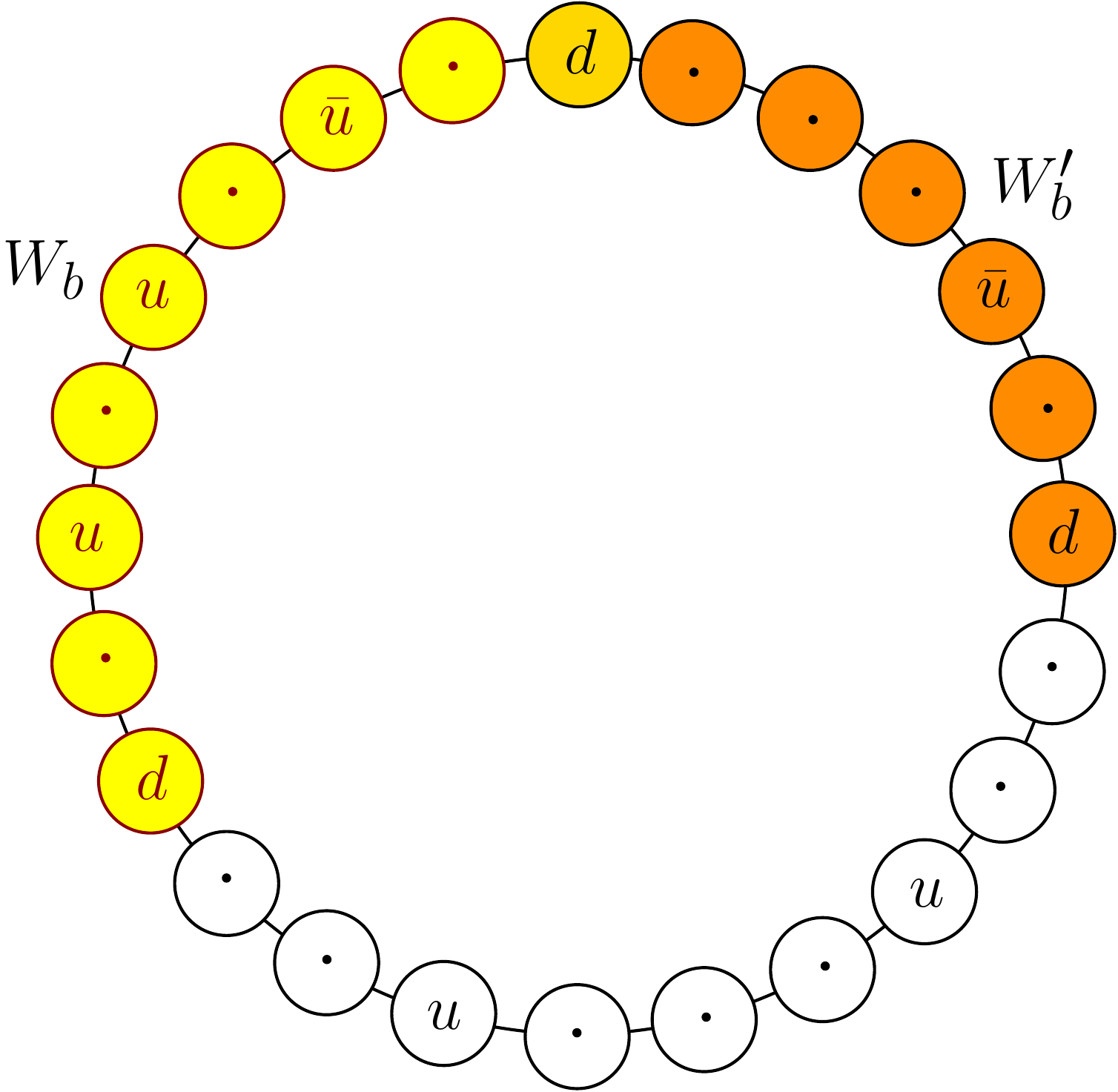}
    \label{fig:vvpn2}
}

\caption[Theorem 2]{Figure showing construction of a solution to the problem with $k+n$ visits, given an optimal solution for the problem with $k$ visits.}
\label{fig:vvpn}
\end{figure}

\IEEEproof
Consider an optimal walk $\mathcal W^*(k)$ with $k \geq $ visits, and any of its binding subwalk, say $\mathcal W_b$. Figure \ref{fig:vvpn1} shows $\mathcal W_b$ (colored yellow). Let $\mathcal W'_b$ be formed by shortcutting all except the last visits to targets from $\mathcal W_b$, such that it has exactly $n$ visits. $\mathcal W'_b$ is shown in Figure \ref{fig:vvpn1} (colored orange). Now form a closed walk $\bar{\mathcal W(k+n)}$ with $k+n$ visits, by inserting $\mathcal W'_b$ in $\mathcal W^*(k)$, right next to $\mathcal W_b$ as shown in Figure \ref{fig:vvpn2}. From Theorem \ref{theorem:concatgeneral}, we have $\mathcal R(\bar{\mathcal W}(k+n)) = \mathcal R(\mathcal W^*(k))$. Since $\mathcal R(\bar{\mathcal W}(k+n))$ is a feasible solution to the problem with $k+n$ visits, the optimal revisit time for a problem with $k+n$ visits is at most $R(\mathcal W^*(k))$, i.e., $\mathcal R(\mathcal W^*(k+n)) \leq \mathcal R(\mathcal W^*(k))$.

\endIEEEproof

 Given an optimal walk with $\mathcal W^*(k)$ visits, one can construct a feasible solution $\bar{\mathcal W}$ with $k+n$ visits such that $\mathcal{ R}(\bar{\mathcal W}) = \mathcal R^*(k)$, by concatenating $\mathcal W^*(k)$ with its shortcut walk of $n$ visits, as discussed above.

\section{Numerical Simulations} \label{sec:sim}
Numerical simulations were performed on 30 instances to corroborate the results proved in the previous section. All the instances contain equally weighted targets, and the number of targets range from 4 to 6. The coordinates of the targets were randomly generated, and the Euclidean distances between them were used as a proxy for the travel times of the UAV. For any instance, an optimal walk can be computed by solving the MILP formulation presented in Appendix \ref{sec:form}, using commercial solvers. \\

\begin{figure}[H]
\centering
\subfigure[Four equally weighted targets]{
    \includegraphics[scale=0.5]{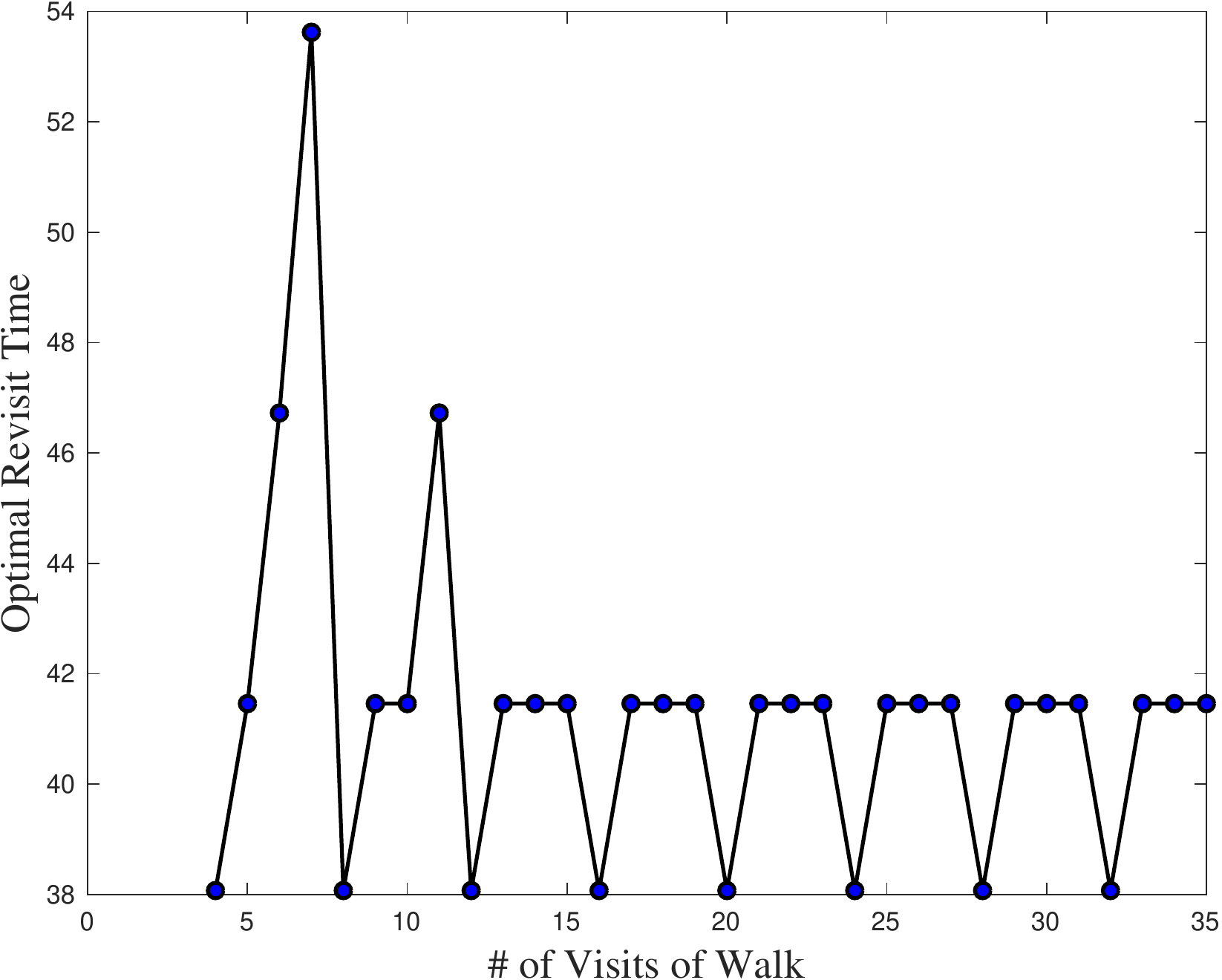}
    \label{fig:equal4}
}
\subfigure[Five equally weighted targets]{
    \includegraphics[scale=0.5]{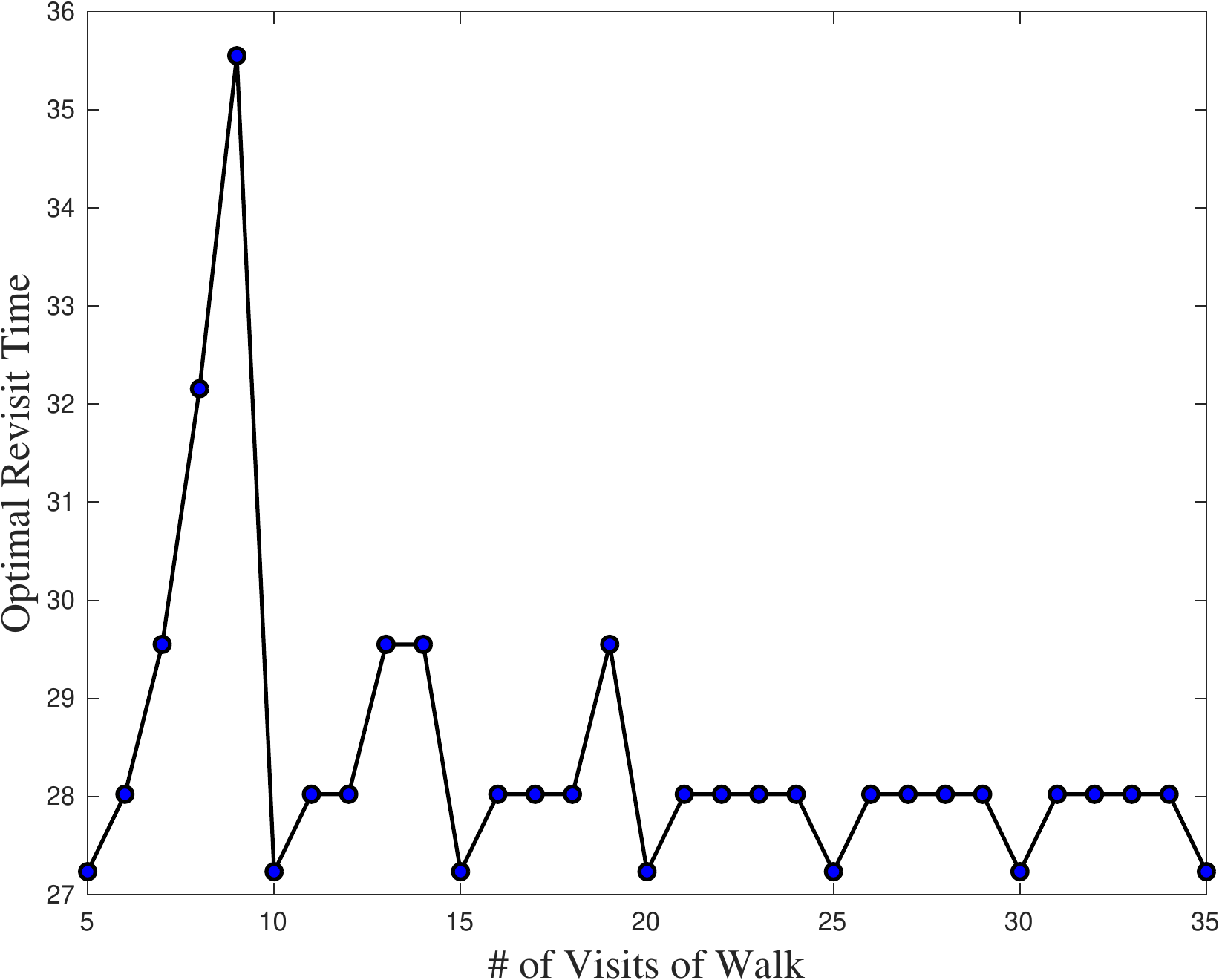}
    \label{fig:equal5}
}
\subfigure[Six equally weighted targets]{
    \includegraphics[scale=0.5]{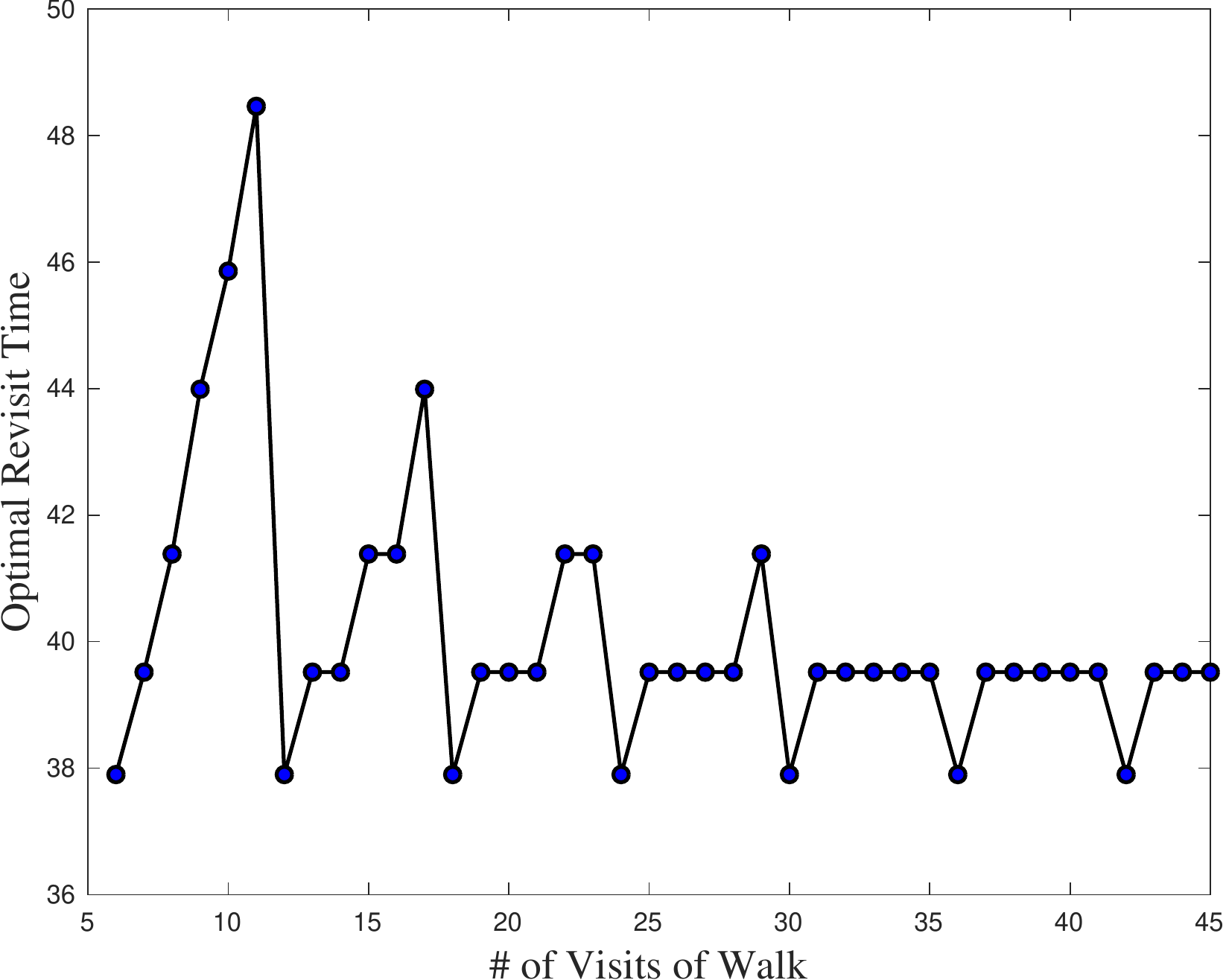}
    \label{fig:equal6}
}
\caption[Equalweights]{The optimal revisit time as a function of the number of visits in the walk.}
\label{fig:equalweights}
\end{figure}

For every instance, optimal solutions were computed using the IBM ILOG CPLEX solver, by varying the allowable number of visits $k$ from $n$ to 45, and their corresponding optimal values were plotted against $k$. Plots for sample instances with 4, 5 and 6 targets are shown in Figures \ref{fig:equal4}, \ref{fig:equal5} and \ref{fig:equal6} respectively.\\

\begin{figure}[H]
    \centering
    \includegraphics[scale=0.5]{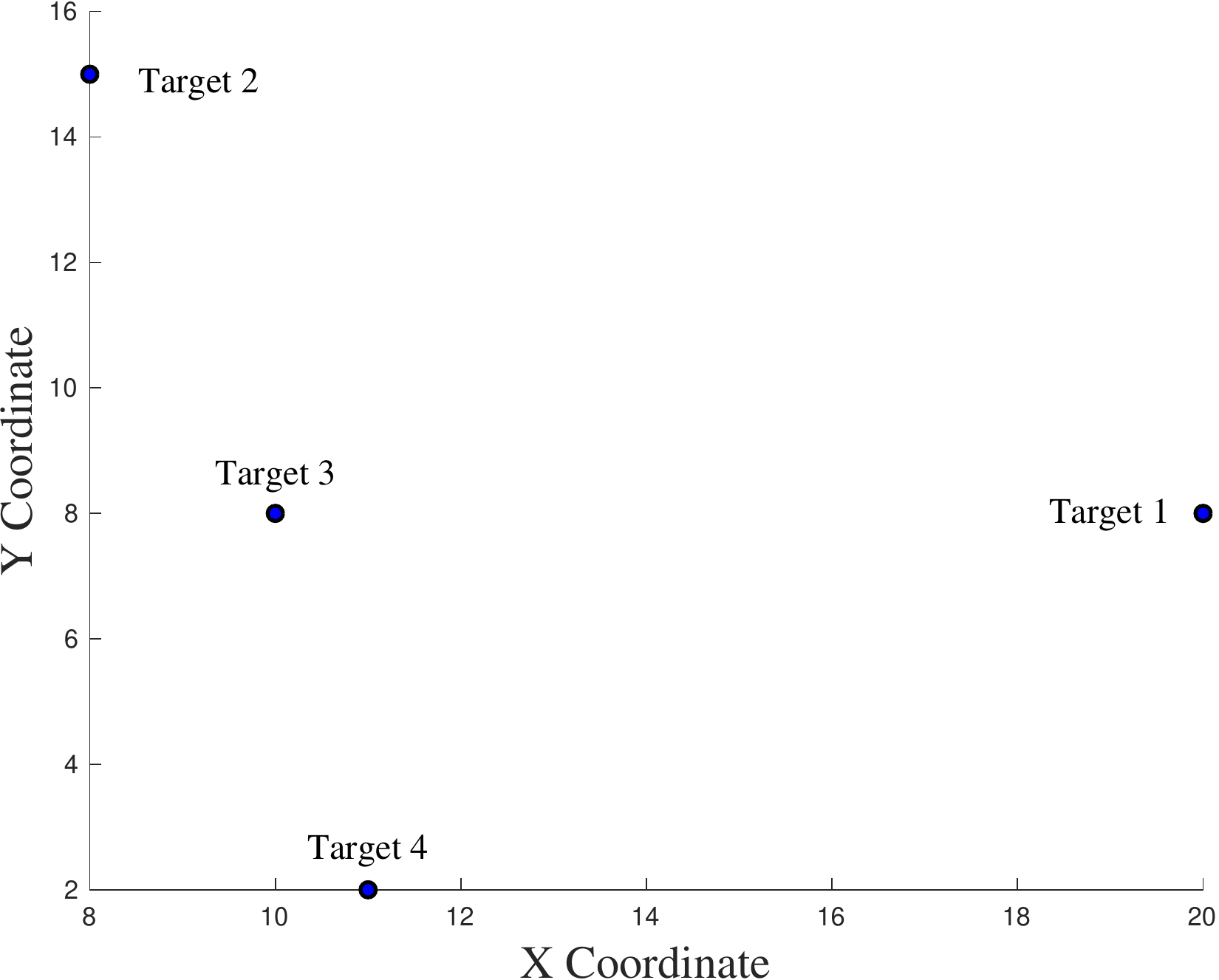}
    \caption{Coordinates of an instance with 4 targets}
    \label{fig:example}
\end{figure}

We consider an instance with 4 targets whose coordinates are shown in Figure \ref{fig:example} to demonstrate the structure of optimal solutions. The time taken by a UAV to travel between the targets is listed in Table \ref{tab:tt}. Optimal solutions are found by varying the number of allowed visits from 4 to 16, and the corresponding optimal solutions are presented in Table \ref{tab:sol}. Figure \ref{fig:equal4} shows the optimal values plotted against the number of visits.

 \begin{table}[h!]
 \centering
    
    \begin{tabular}{c|c|c|c|c|}
    \toprule
     & Target $1$ & Target $2$ & Target $3$ & Target $4$ \\ 
    \midrule
    Target $1$ & 0 & 13.89 & 10 & 10.82\\
    Target $2$ & 13.89 & 0 & 7.28 & 13.34\\
    Target $3$ & 10 & 7.28 & 0 & 6.08\\
    Target $4$ & 10.82 & 13.34 & 6.08 & 0\\
    \bottomrule
    \end{tabular}
    \caption{Travel times between the  targets for the sample instance with 4 targets}
    \label{tab:tt}
\end{table}


For $4 \leq k \leq 7$, there is at least one target that is visited exactly once in a walk (a walk must contain at least $2n$ visits to allow multiple visits to all the targets).

\noindent
So, $\mathcal W^*(k)$ monotonically increases with the increase in number of visits from 4 to 7 as shown in Figure \ref{fig:equal4}. If $k$ is an integral multiple of $n$, the optimal revisit time is equal to $TSP^*$, and an optimal solution can be obtained by repeating TSP solutions.\\

For $k \geq 12$, optimal revisit time takes only two values: $\mathcal R^*(4)$ when $k$ is an integral multiple of $n$, and $\mathcal R^*(5)$ otherwise (Figure \ref{fig:equal4}). In fact any optimal solution $\mathcal W^*(k)$ can be constructed solely using $\mathcal W^*(5)$ and $\mathcal W^*(4)$ when $k \geq 12$. Any $k \geq 12$ can be expressed as $pn+q$ where $p \geq q$, and $p$ and $q$ are integers. When $q = 0$, $k$ is an integral multiple of $n$, and  $\mathcal W^*(k)$ can be obtained by concatenating $\mathcal W^*(4)$ for $p$ times. For example, when $k=12$, $\mathcal W^*(k)$ can be constructed as $\mathcal W^*(4) \circ \mathcal W^*(4) \circ \mathcal W^*(4)$. Since $\mathcal W^*(4) = (2, 3, 4, 1, 2)$, $\mathcal W^*(12)$ is the sequence $(2, 3, 4, 1, 2, 3, 4, 1, 2, 3, 4, 1, 2)$, as shown in Table \ref{tab:sol}. On the other hand, if $q \neq 0$, $\mathcal W^*(k)$ can be constructed by concatenating $\mathcal W^*(n+1)$ for $q$ times, and $\mathcal W(n)$ for $p-q$ times, where $\mathcal W(n)$ is a shortcut walk of $\mathcal W^*(n+1)$, obtained by removing a repeated visit from $\mathcal W^*(n+1)$. For example, consider $k = 14$, which can be expressed in the quotient remainder form as 3(4)+2. Since $p = 3$, $q-p = 2$ and $n+1 = 5$, an optimal solution $\mathcal W^*(14)$ can be obtained by concatenating $\mathcal W^*(5) = (2, 3, 1, 4, 3, 2 )$ for 3 times and its corresponding shortcut walk $\mathcal W(4) = (2, 3, 1, 4, 2)$ for 2 times. That is, $\mathcal W^*(14)$ can be constructed as $\mathcal W^*(5) \circ \mathcal W^*(5) \circ \mathcal W^*(5) \circ \mathcal W(4) \circ \mathcal W(4)$, forming the sequence $(2, 3, 1, 4, 3, 2, 3, 1, 4, 3, 2, 3, 1, 4, 2)$, as one can see from Table \ref{tab:sol}. Therefore, one needs to solve at most 2 problems: $\mathcal W^*(n+1)$ and $\mathcal W^*(n)$ to determine optimal solutions for any $k \geq n(n-1)$.\\

For $4 \leq k < 12$, $\mathcal R^*(k)$ takes at most $n = 4$ distinct values from $\mathcal R^*(4)$ through $\mathcal R^*(7)$ (i.e., $\mathcal R^*(4), \mathcal R^*(5), \mathcal R^*(6), \mathcal R^*(7) $). If $k$ is an integral multiple of $n$, as discussed above, $\mathcal R^*(k) = \mathcal R^*(4)$, and $\mathcal W^*(k)$ can be obtained by repeating $\mathcal W^*(4)$. If $4 \leq k < 12$ is not an integral multiple of $n$, it can be expressed as $pn + q$, such that $p \geq 1$, $1 \leq q \leq n-1$ and $p$ and $q$ are integers. Then, any optimal solution $\mathcal W^*(k)$ can be constructed by concatenating $\mathcal W^*(n+\myceil{\frac{q}{p}})$ and $\mathcal W(n+\floor{\frac{q}{p}})$. For example, consider $k = 11$ which can be expressed in the quotient remainder form as 2(4)+3. Then, $n+\myceil{\frac{q}{p}} = 6$ and $n+\floor{\frac{q}{p}} = 5$. Then, $\mathcal W^*(11) = (2, 3, 1, 3, 4, 3, 2, 3, 1, 3, 4, 2 )$ can be obtained by concatenating $\mathcal W^*(6)$ and $\mathcal W(5)$, where $\mathcal W(5) = ( 2, 3, 1, 3, 4, 2 )$ is obtained by shortcutting a visit from $\mathcal W^*(6) = (2, 3, 1, 3, 4, 3, 2)$. From Table \ref{tab:sol} it can also be observed that $\mathcal R^*(k) \geq \mathcal R^*(k+n)$ for any $k \geq n$, thus corroborating the results proved in Section \ref{sec:proofs}.

 \begin{table}[h!]
 \centering
    
    \begin{tabular}{c|c|c|}
    \toprule
    $k$ & $\mathcal W^*(k)$ & \vtop{\hbox{\strut $\mathcal R^*(k)$}\hbox{\strut  units}}\\
    \midrule
    4 & $(2, 3, 4, 1, 2)$ & 38.07\\
    5 & $(2, 3, 1, 4, 3, 2)$ & 41.46\\
    6 & $(2, 3, 1, 3, 4, 3, 2)$ & 46.73\\
    7 & $(2, 3, 4, 3, 1, 4, 3, 2)$ & 53.63\\
    8 & $(2, 3, 4, 1, 2, 3, 4, 1, 2)$ & 38.07\\
    9 & $(2, 3, 1, 4, 3, 2, 3, 1, 4, 2)$ & 41.46\\
    10 & $(2, 3, 1, 4, 3, 2, 3, 1, 4, 3, 2)$ & 41.46\\
    11 & $(2, 3, 1, 3, 4, 3, 2, 3, 1, 3, 4, 2 )$ & 46.73\\
    12 & $(2, 3, 4, 1, 2, 3, 4, 1, 2, 3, 4, 1, 2)$ & 38.07\\
    13 & $(2, 3, 1, 4, 3, 2, 3, 1, 4, 2, 3, 1, 4, 2)$ & 41.46\\
    14 & $(2, 3, 1, 4, 3, 2, 3, 1, 4, 3, 2, 3, 1, 4, 2)$ & 41.46\\
    15 & $(2, 3, 1, 4, 3, 2, 3, 1, 4, 3, 2, 3, 1, 4, 3, 2)$ & 41.46\\
    16 & $(2, 3, 4, 1, 2, 3, 4, 1, 2, 3, 4, 1, 2, 3, 4, 1, 2 )$ & 38.07\\
    
    \bottomrule
    \end{tabular}
    \caption{Optimal solutions for an instance with 4 targets}
    \label{tab:sol}
\end{table}



\section{Conclusions and Future Work} \label{sec:conc}
In this article we addressed the problem of persistently monitoring a set of $n$ spatially distributed targets of equal priorities, and continuously changing properties, using a UAV. Given $k$ allowable visits as a representative of the vehicle's fuel capacity, the purpose of the work is to identify an optimal walk of $k$ visits such that each target is visited at least once, and the vehicle is refueled at the depot after every $k$ visits. Once the vehicle is refueled, the walk of $k$ visits is continuously repeated, to ensure persistent monitoring of targets. 

In this work, we proved the following:
\begin{enumerate}
    \item $\mathcal R^*(k) \geq TSP^*$, $ \forall  k \geq n$.
    \item $\mathcal R^*(k) = TSP^*$, when $k$ is an integral multiple of $n$.
    \item $\mathcal R^*(k+n) \leq \mathcal R^*(k)$, $\forall k \geq n$.
    \item $\mathcal R^*(k)$ is a monotonically non-decreasing function of $k$, for $n \leq k\leq 2n-1$, .
    \item $\mathcal R^*(k) = \mathcal R^*(n+\myceil{\frac{q}{p}})$, where $k > n$ is not an integral multiple of $n$, and can be expressed as $pn + q$, such that $p \geq 1$, $1 \leq q \leq n-1$ and $p, q \in Z_+$.
    \item $\mathcal R^*(k)$ is a periodic function of $k$, for $k \geq n^2-n$, with a period $n$. Moreover, it takes only two values, $\mathcal R^*(n)$ if $k$ is an integral multiple of $n$, and $\mathcal R^*(n+1)$ otherwise.

\end{enumerate}

A consequence of the results proved in this article is that an optimal walk for any given $k \geq n$ can be constructed using optimal solutions of at most $n$ distinct problems, namely $\mathcal W^*(n)$ to $\mathcal W^*(2n-1)$.

{\it Future work}:  In many practical scenarios, certain targets require higher frequency of monitoring compared to others either due to a much rapid change in properties of these targets, or due to their sheer importance. In such cases, the performance metric is a weighted revisit time, which is the revisit time of a target scaled by its weight. The problem of monitoring weighted targets is a generalization of the problem discussed in this article. Future work will be focused on relaxing the assumption of equally weighted targets, and extending the results presented in this work to a more general setting.
\appendices
\section{Mathematical Formulation} \label{sec:form}

The persistent surveillance problem considered in this work is to identify a walk with certain properties (mentioned above), such that the maximum revisit time among all the targets is a \textit{minimum}, when the targets are visited in the sequence specified by the walk. Mathematical modeling of this problem involves capturing the desired properties of the walk, and the maximum revisit time of all the targets, using a set of constraints. Formulating these constraints and the objective in the form of an MILP is the focus of this section.

Let the binary variable $x^{i,j}_v$ denote the visits made by the vehicle, where $x^{i,j}_v = 1$ if the vehicle travels from target $i$ to target $j$ during its $v^{th}$ visit, and  $x^{i,j}_v = 0$ otherwise. Then, the degree constraints used to model the desired properties of the walk for the problem considered herein are as follows:
\subsection{Degree Constraints}
If the vehicle visits the $i^{th}$ target in its $(v-1)^{th}$ visit, then it must leave the $i^{th}$ target for the $v^{th}$ visit:
 \begin{eqnarray}
 \label{constraint:in=out}\sum_{l = 1}^N x_{v-1}^{li} = \sum_{j = 1}^N x^{ij}_v, \quad \forall i\in \mathcal{T}, \; \; v \in \{2, \dots , k \}. 
 \end{eqnarray}
 Note that the $(v-1)^{th}$ visit and $v^{th}$ visit correspond to different targets in a walk. A vehicle cannot revisit the same target without visiting some other target:
 \begin{align}
 \label{constraint:noselfloop}
  x_v^{ii} = 0, \quad \forall i \in \mathcal{T}, \; \; v \in \{1, \dots , k \}.   
 \end{align} 
Every closed walk begins from and terminates at the same target, identified by the depot ${d}$:
\begin{align}
\label{constraint:depot}
    \sum_{j = 1}^N x^{d,j}_1 = \sum_{l = 1}^N x_{k}^{l,d} = 1.  
\end{align}
Each target must be visited at least once :
\begin{align}
\label{constraint:visitonce}
\sum_{v = 1}^{k} \sum_{j = 1}^{N} x_v^{ij}  \ge 1, \quad \forall i\in \mathcal{T}.
\end{align}
Every visit corresponds to a unique edge (an ordered pair of vertices -- vertex from which the vehicle leaves to the vertex to which vehicle enters).
\begin{align}
    \label{constraint:uniqueedge}
    \sum_{i = 1}^{N} \sum_{j = 1}^{N} x_v^{ij}  = 1, \quad \forall v\in  \{1, \dots , k \}.
\end{align}

These degree constraints force the walk to change with the number of visits because a vehicle is not allowed to stay at the same location (see constraint (\ref{constraint:noselfloop})).

\subsection{Revisit time of targets} \label{subsec:revisit-time}

The revisit time of a target is modeled by bookkeeping the time accumulated since the last visit to the target. The procedure adopted to capture the revisit time of a target involves the following steps:
\begin{enumerate}
    \item After each visit $v$ in the walk, time elapsed since the last visit to a target $i$ is denoted by the variable $f^i_v$.
    \item Unless the $v-1^{th}$ visit of the vehicle is to target $i$, the value $f^i_k$ is higher than $f^i_{v-1}$ by the amount of time required by the vehicle to make its $v^{th}$ visit.
    \item In the case where the $v-1^{th}$ visit in the walk is to target $i$, the time accumulation for target $i$ must restart from zero, and hence $f^i_v$ is equal to the time taken by the vehicle during its $v^{th}$ visit. This is engineered with the help of a variable $z^i_{v-1}$, which equals $f_{v-1}^i$ if the $v-1^{th}$ visit of the vehicle is to target $i$, and zero otherwise (refer to equations \eqref{constraint:zidef} and \eqref{constraint:accumulation}). 
\end{enumerate}
The mathematical way of expressing this scheme is as follows:

\begin{align}
    \label{constraint:zidef}
    z^i_{v-1}  =  f^i_{v-1} \sum_{l= 1}^N x^{li}_{v-1}, \quad \forall i, \; \; v \in  \{2, \dots , k \}.
\end{align}

\begin{align}
\label{constraint:accumulation}
f^i_v - f^i_{v-1} + z^i_{v-1} = \sum_{i = 1}^N \sum_{j = 1}^N x^{ij}_v c(i,j), \; \;\forall i, \; v \in  \{2, \dots , k \}.   
\end{align} 
In equation \eqref{constraint:accumulation}, the purpose of the variable $z^i_{v-1}$ is to negate the time accumulated till the $v-1^{th}$ visit, if the $v-1^{th}$ visit of the vehicle is to target $i$. If the $v-1^{th}$ visit is not to target $i$, then the time accumulated during the $v^{th}$ visit is added to $f^i_{v-1}$.

Since the problem of interest is persistent surveillance, once the vehicle visits all the targets specified by the sequence of the walk, the vehicle keeps repeating the sequence till the end of the mission. Hence, the first visit in the sequence follows the last visit in the previous cycle. Moreover, the $k^{th}$ visit in the sequence is to depot $d$. Hence, by definition,

$$
z^i_{k} =
\begin{cases}
f^i_{k}, i = d,\\
0, i \neq d.
\end{cases}
$$

Consequently, the time accumulation for the first visit in the sequence takes the following form:\\
\begin{align}
    \label{constraint:revisittimparts13}
    f^i_1 - f^i_{k}  = \sum_{j = 1}^N x^{d,j}_1 c(d,j), \quad \forall i \in \mathcal{T} \backslash \{d\}.
\end{align}

\begin{align}
\label{constraint:depotaccumulationtime}
 f^{i}_1 =  \sum_{j = 1}^N x^{d,j}_1 c(d,j), \quad i = d.  
\end{align}


The Mixed Integer Program (MIP) defining the problem can now be posed as follows:
\begin{align}
    \label{obj:nonlinear}
(\mathcal{L}_1):    \min \quad \max_{i\in \mathcal{T}} \{ f_v^i \},
\end{align}
subject to constraints (\ref{constraint:in=out})-(\ref{constraint:uniqueedge}) and (\ref{constraint:zidef})-(\ref{constraint:depotaccumulationtime}). Here the objective function is the maximum revisit time among all the targets, which is $R(W)$ as defined in section \ref{sec:problem-statement}. This formulation can be easily extended to the weighted-case, where the weights $w_i$ associated with the targets are unequal, as follows:
\begin{align}
    \label{obj:nonlinearw}
(\mathcal{L}_2):    \min \quad \max_{i\in \mathcal{T}} \{ w_i f_v^i \},
\end{align}


\subsection{Recasting MIP as MILP}
The objective of MIP can be re-writted in terms of a variable $T$ that acts as a proxy for the maximum revisit time among all the targets, as follows:
\begin{align}
    \label{constraint:obj-linearizing}
     f^i_v  \le T , \quad \forall i \in \mathcal{T}, \; \;  v \in  \{1, \dots , k\}.
\end{align} 
Correspondingly, the objective (\ref{obj:nonlinear}) can be changed to 
\begin{align}
    \label{obj:linear}
   \mathcal R(\mathcal W) = \min T,
\end{align}
subject to constraints  (\ref{constraint:in=out})-(\ref{constraint:uniqueedge}), (\ref{constraint:zidef})-(\ref{constraint:depotaccumulationtime}), and (\ref{constraint:obj-linearizing}).

Note that the constraints (\ref{constraint:zidef}) are bilinear as they involve product of $f_{v-1}^i$ and $x_{v-1}^{li}$. These constraints can be linearized using the standard big-M procedure \cite{griva2009linear}.

After linearization of constraints (\ref{constraint:zidef}), they can be replaced with the following constraints:
\begin{align}
\label{constraint:Mlin1}
 z^i_{v-1} &\le M \sum_{l \in \mathcal{T}} x_{v-1}^{l,i}, \quad \forall i \in \mathcal{T}, \; \;  v \in  \{2, \dots , k \},\\ 
 \label{constraint:Mlin2}
 z^i_{v-1} &\ge f^i_{v-1} - M \left(1 - \sum_{l \in \mathcal{T}}x_{v-1}^{l,i} \right), \forall i \in \mathcal{T}, \; \;  v \in  \{2, \dots , k \},\\
 \label{constraint:Mlin3}
 z^i_{v-1} &\le f^i_{v-1},\quad \forall i \in \mathcal{T}, \; \;  v \in  \{2, \dots , k \},\\ 
 \label{constraint:Mlin4}
 z^i_{v-1} &\ge 0 \quad \forall i \in \mathcal{T}, \; \;  v \in  \{2, \dots , k \}.
\end{align}
The recast MILP can be described with the objective given by (\ref{obj:linear}), subject to linear constraints given by (\ref{constraint:in=out})- (\ref{constraint:uniqueedge}), (\ref{constraint:accumulation})-(\ref{constraint:depotaccumulationtime}), (\ref{constraint:obj-linearizing}), and (\ref{constraint:Mlin1})-(\ref{constraint:Mlin4}).





\section{Additional Proofs} \label{sec:addproofs}

{\it Theorem \ref{theorem:concatgeneral}:
Let $\mathcal W_b$ be a binding subwalk and $S_1$, $S_2$ be subwalks of $\mathcal W$ such that $\mathcal W = S_1 \circ \mathcal W_b \circ S_2$. Let $\mathcal W'_b$ be a closed subwalk obtained by shortcutting visits from $\mathcal W_b$, but retaining the last visits to targets. Let a closed walk $\bar{\mathcal W}$ be formed by concatenating $\mathcal W$ with $\mathcal W'_b$ as follows: $\bar{\mathcal W} =S_1 \circ \mathcal W_b \circ \mathcal W'_b \circ S_2$. Then, $\mathcal R(\bar{\mathcal W}) = \mathcal R(\mathcal W)$.\\ }

\begin{figure}[h!]
\centering

\subfigure[]{
    \includegraphics[scale=0.32]{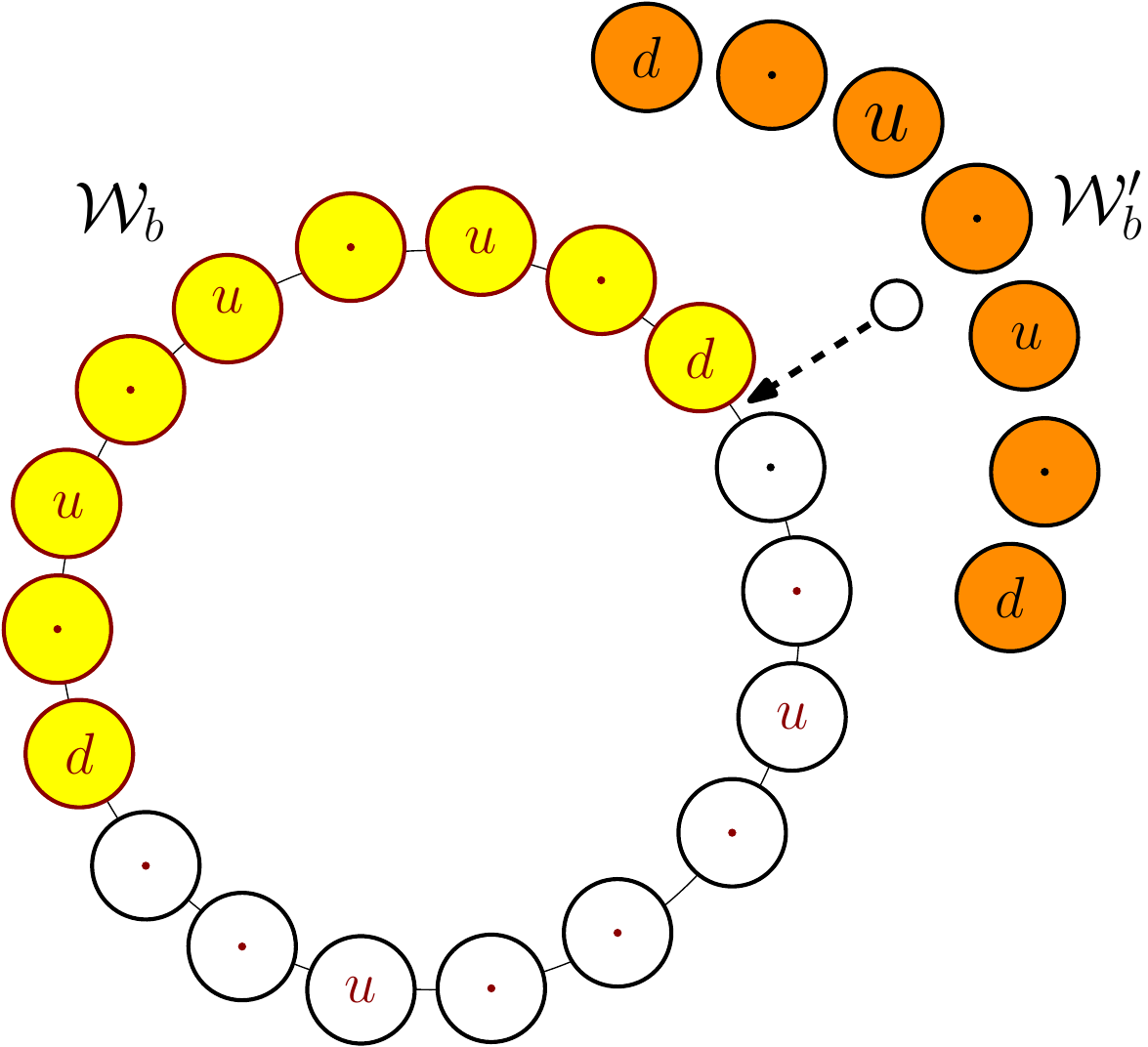}
    \label{fig:concatgen1}
}
\subfigure[]{
    \includegraphics[scale=0.25]{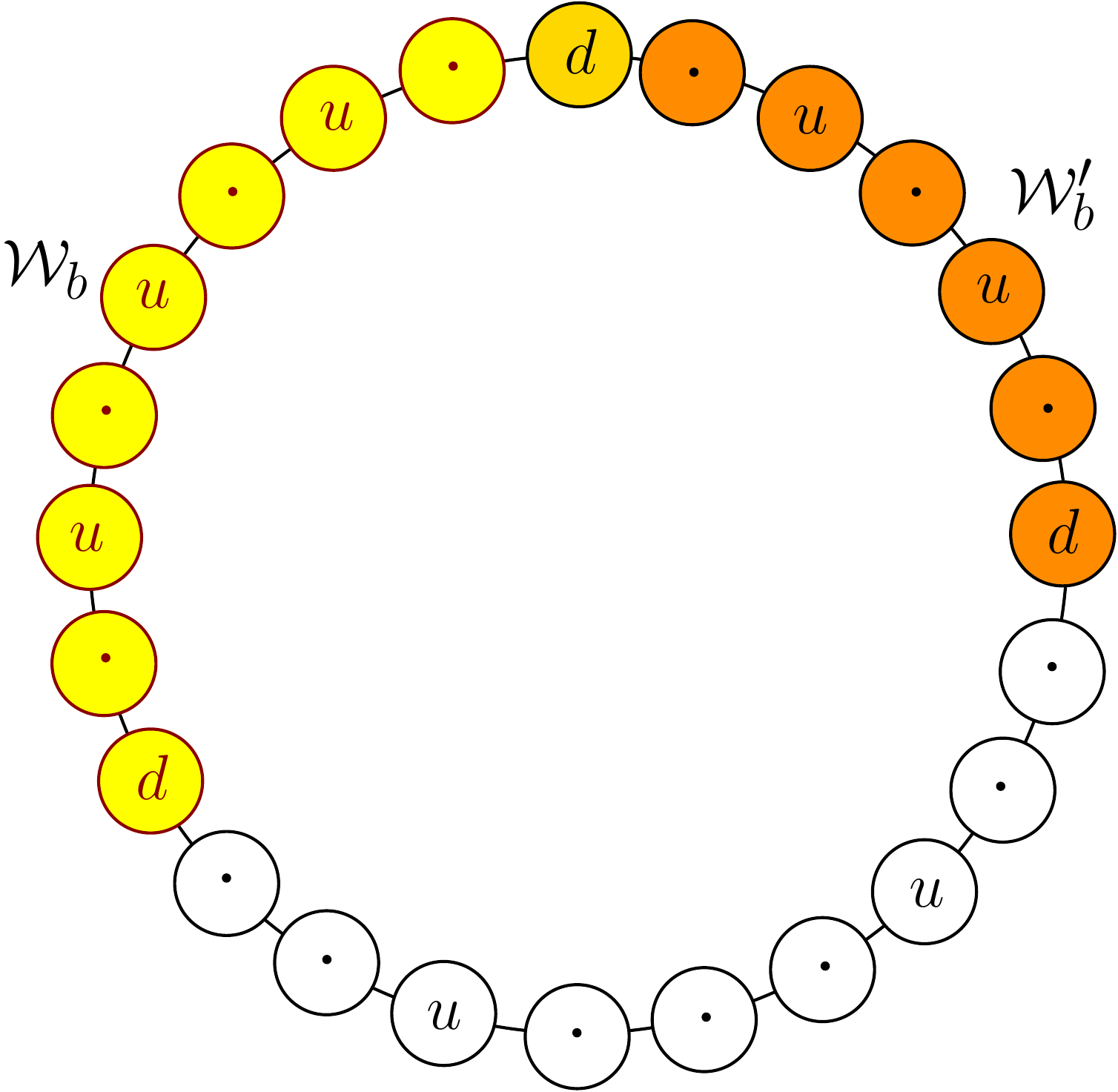}
    \label{fig:concatgen2}
}

\caption[]{A closed walk $\mathcal W$ is shown on the left, with its binding subwalk $\mathcal W_b$ colored yellow. A subwalk $\mathcal W'_b$ (colored orange) is constructed from $\mathcal W_b$, by shortcutting visits to target $u$ as discussed in Theorem \ref{theorem:concatgeneral}. Figure on the right shows the concatenated walk $\bar{\mathcal W}$.}


\label{fig:concatgen}
\end{figure}

 
\IEEEproof
Let target $d$ be the terminus of $\mathcal W_b$; then $d$ is also the terminus of $\mathcal W'_b$, as shown in Figure \ref{fig:concatgen}. By definition, $T(\mathcal W_b) = RT(d, \mathcal W) = \mathcal R(\mathcal W)$. Since $\mathcal W'_b$ is formed by shortcutting visits from $\mathcal W_b$, we have $T(\mathcal W'_b) \leq T(\mathcal W_b) = RT(d, \mathcal W)$. Now consider the closed walk $\bar{\mathcal W} = S_1 \circ \mathcal W_b \circ \mathcal W'_b \circ S_2$. One can see that $RT(d, \bar{\mathcal W}) = \max \{RT(d, \mathcal W), T(\mathcal W'_b)\} = RT(d, \mathcal W) = \mathcal R(\mathcal W)$. 

It is now sufficient to show that for any target $u \neq d$, $RT(u, \bar{\mathcal W}) \leq \mathcal R(\mathcal W)$. Consider a decomposition of $\bar{\mathcal W}$ with respect to $u$. If $u$ is visited $r$ times in $\bar{\mathcal W}$, then there exists a decomposition 
$$ \mathcal C(\bar{\mathcal W},u) = \bar{\mathcal W}_1 \circ \cdots \circ \bar{\mathcal W}_r,$$ 
such that $u$ is the terminus of $\bar{\mathcal W}_1$, \ldots , $\bar{\mathcal W}_r$. According to the above decomposition, the revisit time of $u$ in $\bar{\mathcal W}$ can be expressed as $RT(u, \bar{\mathcal W}) = \max \{T(\bar{\mathcal W}_1), \ldots, T(\bar{\mathcal W}_{r})\}$, and only one of the following cases hold for every $1 \leq i \leq r$:

\begin{enumerate}
    \item $\bar{\mathcal W_i}$ is a closed subwalk of $\mathcal W$,
    \item $\bar{\mathcal W_i}$ is a closed subwalk of $\mathcal W'_b$,
    \item $\bar{\mathcal W_i}$ is a closed subwalk of $\mathcal W \circ \mathcal W'_b$, but is not a subwalk of either $\mathcal W$ or $\mathcal{W}_b'$,
    \item $\bar{\mathcal W_i}$ is a closed subwalk of $\mathcal W'_b \circ \mathcal W$, but is not a subwalk of either $\mathcal W$ or $\mathcal{W}_b'$.\\
\end{enumerate}

In case 1, by definition, $T(\bar{\mathcal W}_i) \leq \mathcal R(\mathcal W)$. Similarly, in case 2, we have $T(\bar{\mathcal W}_i) \leq T(\bar{\mathcal W'}_b) \leq T(\bar{\mathcal W}_b) = \mathcal R(\mathcal W)$.
To show $T(\bar{\mathcal W}_i) \leq  \mathcal R(\mathcal W)$ for cases 3 and 4, we consider decompositions of $\mathcal W_b$ and $\mathcal W'_b$ with respect to their last visists to $u$ as discussed below.\\

$\mathcal W_b$ is decomposed with respect to its last visit to $u$ as:
$$ \mathcal W_b = \mathcal S_{b_1} \circ \mathcal S_{b_2},$$
such that the last visit to $u$ in $\mathcal W_b$ appears as the last node of $\mathcal S_{b_1}$ and the first node of $\mathcal S_{b_2}$, and $d$ is the first node of the subwalk $\mathcal S_{b_1}$, and the last node of the subwalk $\mathcal S_{b_2}$.\\

Similarly, $\mathcal W'_b$ is decomposed with respect to the last visit to $u$ as:
$$ \mathcal W'_b =  \hat{\mathcal S}_{b_1} \circ \hat{\mathcal S}_{b_2},$$
such that the last visit to $u$ in $\mathcal W'_b$ appears as the the last node of $\hat{\mathcal S}_{b_1}$, but the first node of $\hat{\mathcal S}_{b_1}$, and $d$ is the first node of $\hat{\mathcal S}_{b_1}$ and the last node of $\hat{\mathcal S}_{b_2}$.

Since $\mathcal W'_b$ is formed by shortcutting visits from $\mathcal W_b$, but retaining the last visit to $u$, we have $T(\hat{\mathcal S}_{b_1}) \leq T(\mathcal S_{b_1})$, and $T(\hat{\mathcal S}_{b_2}) \leq T(\mathcal S_{b_2})$. \\

\begin{figure}[h!]
    \centering
    \includegraphics[scale=0.3]{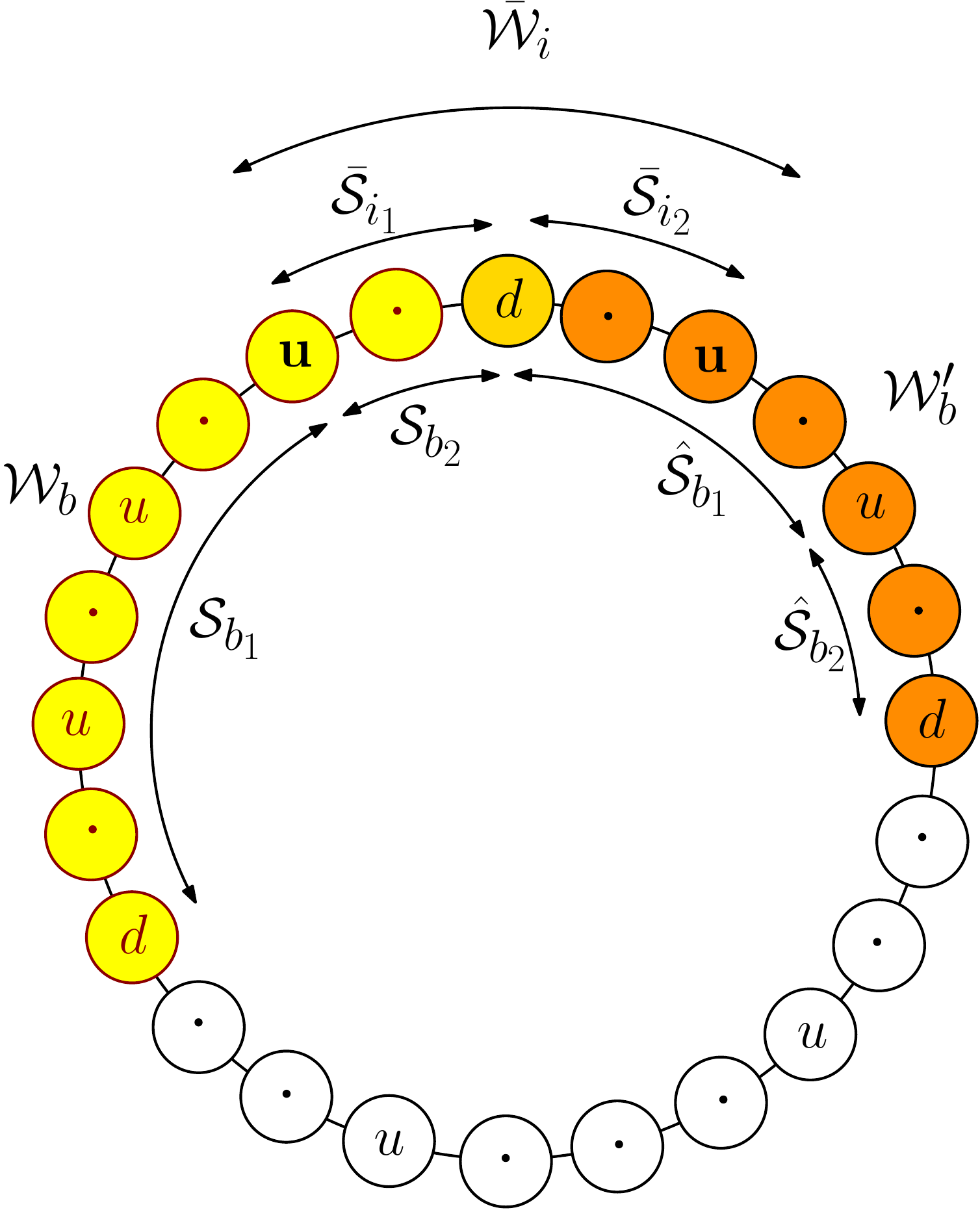}
    \caption{A closed walk $\bar{\mathcal W}$, and its corresponding subwalks as in case 3 of Theorem \ref{theorem:concatgeneral}.}
    \label{fig:concatgen2_3}
\end{figure}

Consider case 3; $\bar{\mathcal W}_i$ has its initial node in $\mathcal W$, and terminal node in $\mathcal W'_b$. By construction, $\mathcal W_b$ precedes $\mathcal W'_b$ in $\bar{\mathcal W}$. Moreover, from Lemma \ref{lemma:binding}, $u$ is visited in $\mathcal W_b$. Therefore, the first node of  $\bar{\mathcal W}_i$ is in $\mathcal W_b$, and the last node of $\bar{\mathcal W}_i$ is in $\mathcal W'_b$. So, as shown in Figure  \ref{fig:concatgen2_3}, $\bar{\mathcal W}_i$ can be decomposed with respect to $d$ as:
$$\bar{\mathcal W}_i = \bar{\mathcal S}_{i_1} \circ \bar{\mathcal S}_{i_2}, $$
where $\bar{\mathcal S}_{i_1}$ is a subwalk of $\mathcal W_b$ with $u$ and $d$ as its first and last nodes respectively, and $\bar{\mathcal S}_{i_2}$ is a subwalk of $\mathcal W'_b$ with $d$ and $u$ as its first and last nodes respectively.

By construction, $\bar{\mathcal S}_{i_1} = \mathcal S_{b_2}$, and $\bar{\mathcal S}_{i_2}$ is a subwalk of $\hat{\mathcal S}_{b_1}$. Thereby, we have $T(\bar{\mathcal S}_{i_1}) = T(\mathcal S_{b_2})$, and $T(\bar{\mathcal S}_{i_2}) \leq T(\hat{\mathcal S}_{b_1})$ (note that equality holds when $\mathcal W'_b$ is formed by shortcutting all visits to $u$ except for the last visit to it). Thereby, it follows that $T(\bar{\mathcal W_i}) = T(\bar{\mathcal S}_{i_1}) + T(\bar{\mathcal S}_{i_2}) =  T(\mathcal S_{b_2}) + T(\hat{\mathcal S}_{b_1}) \leq  T(\mathcal S_{b_2}) + T(\mathcal S_{b_1})  = T(\mathcal W_b) = \mathcal R(\mathcal W)$.\\

\begin{figure}[h!]
    \centering
    \includegraphics[scale=0.3]{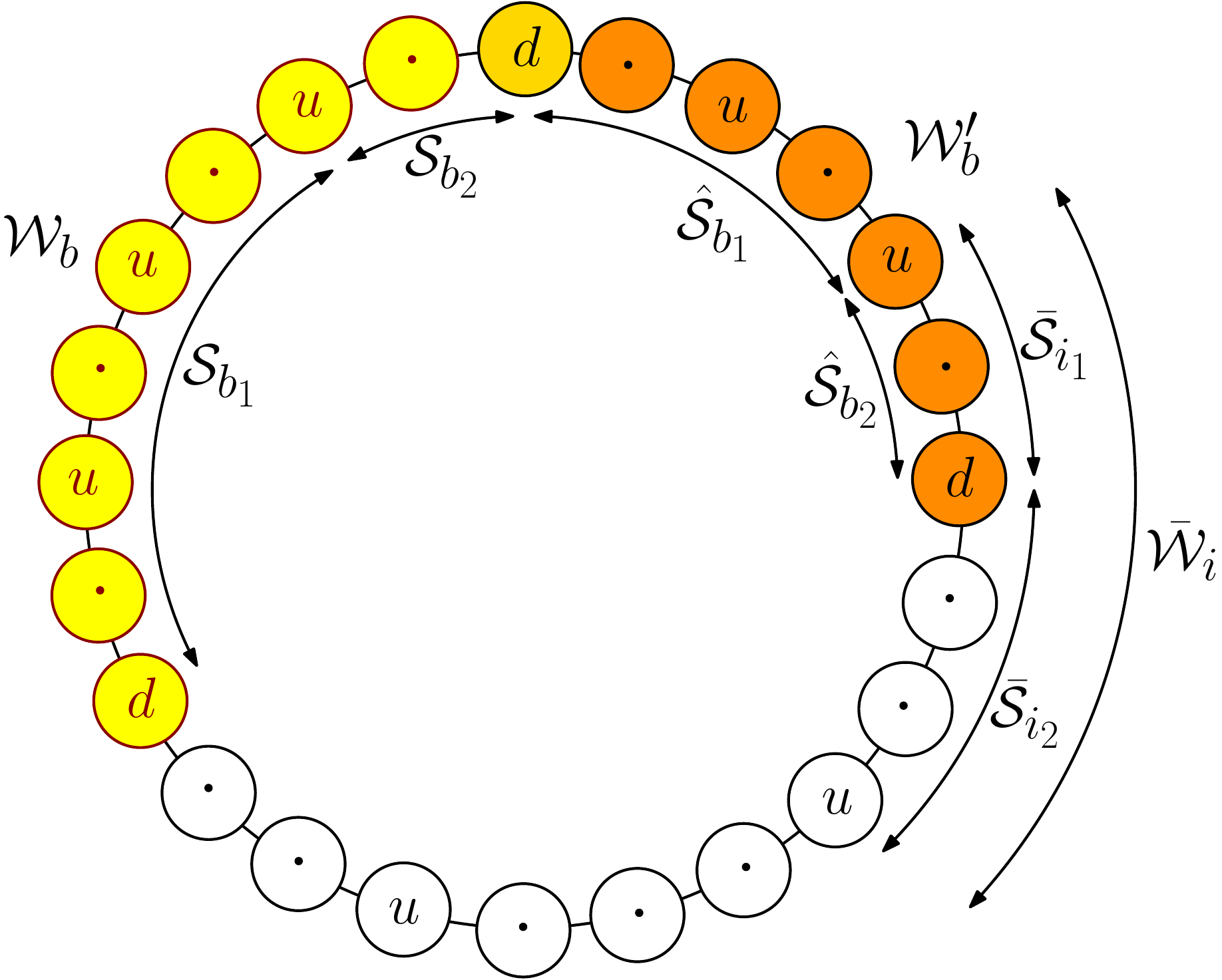}
    \caption{A closed walk $\bar{\mathcal W}$, and its corresponding subwalks as in case 4 of Theorem \ref{theorem:concatgeneral}.}
    
    
    \label{fig:concatgen2_4}
\end{figure}

Consider case 4; the first node of $\bar{\mathcal W}_i$ is in $\mathcal W'_b$, and the last node of $\bar{\mathcal W}_i$ is in $\mathcal W$. Then, as shown in Figure \ref{fig:concatgen2_4}, $\bar{\mathcal W}_i$ can be decomposed with respect to $d$ as: 
$$\bar{\mathcal W}_i = \bar{\mathcal S}_{i_1} \circ \bar{\mathcal S}_{i_2}, $$
where $\bar{\mathcal S}_{i_1}$ is a subwalk of $\mathcal W'_b$ with $u$ and $d$ as its first and last nodes respectively, and $\bar{\mathcal S}_{i_2}$ is a subwalk of $\mathcal W$ (or $\mathcal S_2$) with $d$ and $u$ as its first and last nodes respectively.

By construction, $\bar{\mathcal S}_{i_1} = \hat{\mathcal S}_{b_2}$ and $T(\bar{\mathcal S}_{i_1}) = T(\hat{\mathcal S}_{b_2})$. Therefore, $T(\bar{\mathcal W_i}) = T(\bar{\mathcal S}_{i_1}) + T(\bar{\mathcal S}_{i_2}) = T(\hat{\mathcal S}_{b_2}) + T(\bar{\mathcal S}_{i_2}) \leq T(\mathcal S_{b_2}) + T(\bar{\mathcal S}_{i_2}) \leq RT(u, \mathcal W) \leq \mathcal R(\mathcal W)$. Note that the second inequality follows from the fact that $\mathcal S_{b_1} \circ \bar{\mathcal S}_{i_2}$ is a closed subwalk of $\mathcal W$ with $u$ as its terminus.

Hence, we proved that for any arbitrary node $u \neq d$, $RT(u, \bar{\mathcal W}) \leq \mathcal R(\mathcal W)$, and $RT(d, \bar{\mathcal W} = \mathcal R(\mathcal W)$. So, by definition, $\mathcal R(\bar{\mathcal W}) = RT(d, \bar{\mathcal W} = \mathcal R(\mathcal W)$.

Therefore, we proved that $RT(u,\bar{\mathcal W}) \leq \mathcal R(\mathcal W)$. Since $u$ is arbitraty, this is true for all the targets, and we have, $\mathcal R(\bar{\mathcal W}) \leq \mathcal R(\mathcal W)$. This together with $\mathcal R(\bar{\mathcal W}) \leq \mathcal R(\mathcal W)$ proved above implies $\mathcal R(\bar{\mathcal W}) = \mathcal R(\mathcal W)$.

\endIEEEproof

\noindent
{\it Lemma \ref{lemma:sameconcat}:
 Let a closed walk $\bar{\mathcal W}$ be formed by concatenating $\mathcal W(k)$ with itself $p$ times as follows:
     $$\bar{\mathcal W} :=  \underbrace{\mathcal W(k)\circ \cdots \circ \mathcal W(k)}_{p ~times}. $$
Then, for any node $v_r$
\begin{itemize}
    \item 
    $\mathcal{C}(\bar{\mathcal{W}},v_r) = \underbrace{\mathcal{C}(\mathcal{W}(k),v_r) \circ 
    \cdots \circ \mathcal{C}(\mathcal{W}(k),v_r)}_{p \; times}.$
    
    \item $RT(v_r,\bar{\mathcal W}) = RT(v_r,\mathcal W(k))$, 
\end{itemize}
and thus ${\mathcal R}(\bar{\mathcal W}) = \mathcal R (\mathcal W(k)).$\\
  }

\IEEEproof Let $\mathcal W = (v_1, v_2, \ldots v_k, v_1)$ with $v_r$ being an intermediate node. Then,
$$\bar{\mathcal{W}} = (\underbrace{v_1, v_2, \ldots, v_k}_{1},
\underbrace{v_1, v_2, \ldots, v_k}_{2}, 
\ldots, \underbrace{v_1, v_2, \ldots, v_k}_{p}, v_1). $$
Clearly, then
\begin{align*}
    \mathcal{C}(\bar{\mathcal{W}}, v_r) = (\underbrace{v_r, v_{r+1}, \ldots, v_k, v_1, v_2, \ldots, v_{r-1}}_{1}, \\
    \underbrace{v_r, v_{r+1}, \ldots, v_k, v_1, v_2, \ldots, v_{r-1}}_{2},\\
    \vdots \hspace{6.8em}\\
    \underbrace{v_r, v_{r+1}, \ldots, v_k, v_1, v_2, \ldots, v_{r-1}}_{p}, v_r)\\
    = \underbrace{\mathcal{C}(\mathcal{W},v_r) \circ \mathcal{C}(\mathcal{W},v_r)\circ 
    \cdots \circ \mathcal{C}(\mathcal{W},v_r)}_{p \; times}
\end{align*}

Let $v_r$ be visited $l$ times in $\mathcal W$. Then there exist closed subwalks $\mathcal W_1, \mathcal W_2, \ldots \mathcal W_l $ with $v_r$ as a terminus such that 
$$ \mathcal C(\mathcal W(k),v_r) = \mathcal W_1 \circ \mathcal W_2 \circ \ldots \circ \mathcal W_l. $$

Hence, 
\begin{align*}
 \mathcal C(\bar{\mathcal W},v_r) = \underbrace{\mathcal W_1 \circ \mathcal W_2 \circ \ldots \circ \mathcal W_l}_{1} \circ\\
 \quad \underbrace{\mathcal W_1 \circ \mathcal W_2 \circ \ldots \circ \mathcal W_l}_{2} \circ\\
 \vdots \hspace{4.7em}\\
 \underbrace{\mathcal W_1 \circ \mathcal W_2 \circ \ldots \circ \mathcal W_l}_{p}.
\end{align*}

By definition, we have 
$$RT(v_r,\mathcal{C}(\bar{\mathcal W},v_r)) = \max_{1 \le i \le l}  T(\mathcal W_i) \text{, and} $$
$$RT(v_r,\mathcal{C}(\mathcal W(k),v_r)) = \max_{1 \le i \le l}  T(\mathcal W_i).$$
$$ \implies RT(v_r,\mathcal{C}(\bar{\mathcal W},v_r)) = RT(v_r,\mathcal{C}(\mathcal W(k),v_r)), \forall v_r \in \mathcal{T} .$$
Hence, by definition,
 $RT(v_r,\bar{\mathcal W}) = RT(v_r,\mathcal{C}(\bar{\mathcal W},v_r)) = 
    RT(v_r,\mathcal{C}(\mathcal W(k),v_r)) = RT(v_r,\mathcal W(k)), \quad \forall v_r \in \mathcal{T}$.

Therefore,
$$ \mathcal R(\bar{\mathcal W}) = 
\mathcal R(\mathcal W(k)).$$


\endIEEEproof


\noindent
{\it Lemma \ref{lemma:permut}:
Let $\mathcal W(k)$ be a closed walk with $n+1 \le k \le 2n-1$ visits, and $d$ be a target that is visited exactly once in $\mathcal W(k)$. Let $\mathcal W(k-1)$ be a walk formed by shortcutting a revisit from  $\mathcal W(k)$. Consider a closed walk $\bar{\mathcal W}$ formed by concatenating $\mathcal W(k)$ for $q~ (\geq 1)$ times and $\mathcal W(k-1)$ for $p$ times as follows:
      $$\bar{\mathcal W} := \underbrace{\mathcal W(k)\circ ..\circ \mathcal W(k)}_{q ~times}\circ \underbrace{\mathcal W(k-1)\circ .. \circ \mathcal W(k-1)}_{p ~times}, $$
Then, 
\begin{multline*}
\mathcal{C}(\bar{\mathcal{W}},d) = 
\underbrace{\mathcal{C}(\mathcal{W}(k),d) \circ \cdots \circ \mathcal{C}(\mathcal{W}(k),d)}_{q \; times} \circ \\ \underbrace{\mathcal{C}(\mathcal{W}(k-1),d) \circ \cdots \circ \mathcal{C}(\mathcal{W}(k-1),d)}_{p \; times}. 
\end{multline*}
  }

\IEEEproof
Let $\mathcal W(k) = (v_1, v_2, \ldots v_r \ldots v_k, v_1)$, with $v_1$ as the initial node and let $v_r = d$. We may construct $\mathcal W(k-1)$  by shortcutting a node $v_s$ from $\mathcal W(k)$, where $ 1 \leq s (\neq r) \leq k$.\\
Consider the case $s > r$. Then,
\begin{multline*}
     \mathcal W(k) = (v_1, v_2, \ldots, v_{r-1}, d, v_{r+1}, \ldots, v_s \ldots v_k, v_1),\\
     \text{and }\mathcal W(k-1) = (v_1, v_2, \ldots, v_{r-1}, d, v_{r+1}, \ldots \\,v_{s-1}, v_{s+1}, \ldots, v_k, v_1).
\end{multline*}

Their cyclic permutations with $d$ as the terminus are:
$$ \mathcal C(\mathcal W(k),d) = (d, v_{r+1},  \ldots, v_s \ldots v_k, v_1, \ldots, v_{r-1}, d), $$
\begin{multline*}
  \text{and } \mathcal C(\mathcal W(k-1),d) = \\
  (d, v_{r+1}, \ldots, v_{s-1}, v_{s+1},\ldots, v_k, v_1, \ldots, v_{r+1}, d). 
\end{multline*}

Then,
\begin{multline*}
     \underbrace{\mathcal{C}(\mathcal{W}(k),d) \circ \cdots \circ \mathcal{C}(\mathcal{W}(k),d)}_{q \; times} \circ   \\     \underbrace{\mathcal{C}(\mathcal{W}(k-1),d) \circ \cdots \circ \mathcal{C}(\mathcal{W}(k-1),d)}_{p \; times} = 
\end{multline*}
\begin{align*}
    & \underbrace{(d, v_{r+1},  \ldots, v_s, \ldots, v_k, v_1, \ldots, v_{r-1},}_{1}\\
    & \vdotswithin{\underbrace{(d, v_{r+1},  \ldots, v_s, \ldots, v_k, v_1, \ldots, v_{r-1},}_{1}} \\
    & \underbrace{d, v_{r+1},  \ldots, v_s \ldots v_k, v_1, \ldots, v_{r-1},}_{q}\\
    & \underbrace{d, v_{r+1}, \ldots, v_{s-1}, v_{s+1},\ldots, v_k, v_1, \ldots, v_{r-1},}_{1} \\
    & \vdotswithin{\underbrace{d, v_{r+1}, \ldots, v_{s-1}, v_{s+1},\ldots, v_k, v_1, \ldots, v_{r-1},}_{1}}\\
    & \underbrace{d, v_{r+1}, \ldots, v_{s-1}, v_{s+1},\ldots, v_k, v_1, \ldots, v_{r-1}, d).}_{p}
\end{align*}

Now consider
\begin{multline*}
      \bar{\mathcal W} := \underbrace{\mathcal W(k)\circ ..\circ \mathcal W^*(k)}_{q ~times}\circ \underbrace{\mathcal W(k-1)\circ .. \circ \mathcal W(k-1)}_{p ~times} =  \\
\end{multline*}
\begin{align*}
    & \underbrace{(v_1, v_2, \ldots, v_{r-1}, \circled{d}, v_{r+1}, \ldots v_s \ldots v_k,}_{1} \\
    & \vdotswithin{\underbrace{(v_1, v_2, \ldots, v_{r-1}, \circled{d}, v_{r+1}, \ldots v_s \ldots v_k,}_{1} }\\
    & \underbrace{(v_1, v_2, \ldots, v_{r-1}, d, v_{r+1}, \ldots v_s \ldots v_k,v_1)}_{p}\\
    & \underbrace{(v_1, v_2, \ldots, v_{r-1}, d, v_{r+1}, \ldots v_{s-1}, v_{s+1} \ldots v_k,)}_{1}\\
    & \vdotswithin{\underbrace{(v_1, v_2, \ldots, v_{r-1}, d, v_{r+1}, \ldots v_{s-1}, v_{s+1} \ldots v_k,)}_{1}}\\
    & \underbrace{v_1, v_2, \ldots, v_{r-1}, \bm{d}, v_{r+1}, \ldots v_{s-1}, v_{s+1} \ldots v_k, v_1)}_{q}.
\end{align*}

A cyclic permutation of $\bar{\mathcal W}$, with its first visit to $d$ (circled) as the initial node can be written as, 
\begin{align*}
   {\mathcal{C}(\bar{\mathcal{W}},d)} = & \notag  \\
    & \underbrace{(d, v_{r+1},  \ldots, v_s \ldots v_k, v_1, \ldots, v_{r-1},}_{1} \\
    & \vdotswithin{\underbrace{(d, v_{r+1},  \ldots, v_s \ldots v_k, v_1, \ldots, v_{r-1},}_{1}} \\
    & \underbrace{d, v_{r+1},  \ldots, v_s \ldots v_k, v_1, \ldots, v_{r-1},}_{q} \\
    & \underbrace{d, v_{r+1}, \ldots, v_{s-1}, v_{s+1},\ldots, v_k, v_1, \ldots, v_{r-1},}_{1} \\
    &  \vdotswithin{\underbrace{d, v_{r+1}, \ldots, v_{s-1}, v_{s+1},\ldots, v_k, v_1, \ldots, v_{r-1},}_{1}} \\
    & \underbrace{d, v_{r+1}, \ldots, v_{s-1}, v_{s+1},\ldots, v_k, v_1, \ldots, v_{r-1}, d).}_{p}
\end{align*}

Hence, it can be seen that 
\begin{multline*}
\mathcal{C}(\bar{\mathcal{W}},d) = 
\underbrace{\mathcal{C}(\mathcal{W}(k),d) \circ \cdots \circ \mathcal{C}(\mathcal{W}(k),d)}_{q \; times} \circ \\ \underbrace{\mathcal{C}(\mathcal{W}(k-1),d) \circ \cdots \circ \mathcal{C}(\mathcal{W}(k-1),d)}_{p \; times}. 
\end{multline*}

The same can be shown for the case $s < r$, by considering a permutation of $\bar{\mathcal W}$ with the last visit to $d$ (shown in bold) as its initial node.

\endIEEEproof

\bibliographystyle{IEEEtran}
\bibliography{references.bib}

\end{document}